\documentclass[twocolumn]{aastex62}

\usepackage{color}
\usepackage{amsmath}
\usepackage{enumitem}
\usepackage{graphicx}
\usepackage{mathrsfs}
\usepackage{booktabs}
\usepackage{morefloats}
\usepackage{float}
\usepackage{placeins}

\newcommand{\code}[1]{\texttt{\detokenize{#1}}}

\newcommand{\thecannon}{\textsl{The Cannon}}
\newcommand{\cannon}{\textsl{Cannon}}
\newcommand{\gaia}{\textsl{Gaia}}
\newcommand{\drtwo}{\textsl{Gaia-DR2}}

\newcommand{\kepler}{\textsl{Kepler}}

\newcommand{\tess}{\textsl{TESS}}
\newcommand{\msun}{M_{\odot}}
\newcommand{\teff}{T_{\mathrm{eff}}}
\newcommand{\logg}{\log g}
\newcommand{\feh}{[{\mathrm{Fe}/\mathrm{H}}]}
\newcommand{\mh}{[{\mathrm{M}/\mathrm{H}}]}

\definecolor{bcolor}{RGB}{0, 51, 153}
\definecolor{gcolor}{RGB}{51, 153, 51}
\hypersetup{linkcolor=bcolor,citecolor=bcolor,filecolor=cyan,urlcolor=bcolor}

\received{August 23, 2019}
\revised{January 13, 2020}
\accepted{January 15, 2020}
\submitjournal{\apj}

\shorttitle{temperatures and metallicities for m dwarfs}
\shortauthors{birky et al.}

\begin{document}\raggedbottom\sloppy\sloppypar\frenchspacing

\title{TEMPERATURES AND METALLICITIES OF M DWARFS IN THE APOGEE SURVEY}

\correspondingauthor{Jessica Birky}
\email{jbirky@uw.edu}

\author[0000-0002-7961-6881]{Jessica Birky}
\affil{Center for Astrophysics and Space Science, University of California San Diego, La Jolla, CA 92093, USA}
\affil{Max-Planck-Institut f\"ur Astronomie, K\"onigstuhl 17, D-69117 Heidelberg, Germany}

\author[0000-0003-2866-9403]{David W. Hogg}
\affil{Max-Planck-Institut f\"ur Astronomie, K\"onigstuhl 17, D-69117 Heidelberg, Germany}
\affil{Center for Cosmology and Particle Physics, Department of Physics, New York University, 726
Broadway, New York, NY 10003, USA}
\affil{Center for Data Science, New York University, 60 Fifth Ave, New York, NY 10011, USA}
\affil{Center for Computational Astrophysics, Flatiron Institute, 162 Fifth Ave, New York, NY 10010, USA}

\author[0000-0003-3654-1602]{Andrew W. Mann}
\affil{Department of Astronomy, Columbia University, 550 West 120th Street, New York, NY 10027, USA}
\affil{Department of Physics and Astronomy, The University of North Carolina at Chapel Hill, Chapel Hill, NC 27599, USA}

\author[0000-0002-6523-9536]{Adam Burgasser}
\affil{Center for Astrophysics and Space Science, University of California San Diego, La Jolla, CA 92093, USA}

\begin{abstract}

M dwarfs have enormous potential for our understanding of structure and
formation on both Galactic and exoplanetary scales
through their properties and compositions.
However, current atmosphere models
have limited ability to reproduce spectral features in stars at the 
coolest temperatures ($T_{\rm eff} < 4200\,$K) and to fully
exploit the information content of current and upcoming large-scale spectroscopic surveys.
Here we present a catalog of spectroscopic temperatures, metallicities, and
spectral types for 5875 M dwarfs in the Apache Point Observatory
Galactic Evolution Experiment (APOGEE) and \drtwo\ surveys using
\thecannon: a flexible, data-driven spectral-modeling and
parameter-inference framework demonstrated to estimate
stellar-parameter labels ($\teff$, $\logg$, $\feh$, and detailed
abundances) to high precision.
Using a training sample of 87 M dwarfs with optically derived labels spanning 
$2860 < \teff < 4130$\,K calibrated with bolometric temperatures, and
$-0.5 < \feh < 0.5$\,dex calibrated with FGK binary metallicities, 
we train a two-parameter model with predictive accuracy (in cross-validation) 
to 77\,K and 0.09\,dex respectively.
We also train a one-dimensional spectral classification model using 51 M dwarfs
with Sloan Digital Sky Survey optical spectral types ranging from M0 to M6, to predictive accuracy of
0.7 types. 
We find \cannon\ temperatures to be in agreement to within 60\,K
compared to a subsample of 1702 sources with color-derived temperatures, and 
\cannon\ metallicities to be in agreement to within 0.08\,dex metallicity
compared to a subsample of 15 FGK+M or M+M binaries. 
Finally, our comparison between \cannon\ and APOGEE pipeline (ASPCAP DR14) labels finds that 
ASPCAP is systematically biased toward reporting higher temperatures and lower metallicities 
for M dwarfs. 
\end{abstract}


\section{Introduction} \label{sec:intro}

Low-mass stars, with masses $M_{*} < 0.7\,\msun$ and effective temperatures $T_{\rm eff} < 4000\,$K, are by far the most ubiquitous type of star, comprising $\sim 70 \%$ of the Galaxy's population by number \citep{Bochanski:2010}. With nuclear fusion timescales $\tau>10^{11}$ yr \citep{Laughlin:1997}, the chemical compositions of the M-dwarf population trace the nucleosynthetic processes and interstellar mixing of heavy elements from many generations of shorter-lived, high-mass stars, and are a unique probe for piecing together Galactic structure and evolution \citep{Bochanski:2010,Woolf:2012}.  

Additionally, the low masses of M dwarfs make for easier
detection of planets by variability in radial velocity \citep{Trifonov:2018}, 
high ratios of planet-to-star radii make for easier
detection of exoplanet transits in observations of light curves
\citep{Nutzman:2008}, and shorter orbital periods
(for a fixed stellar insolation flux) allow for discovery of new
planets in less observation time than for more massive stars. 
For these reasons, M dwarfs are primary candidates for exoplanet searches, including by the NASA \kepler\ \citep[e.g.,][]{Dressing2015} and \textsl{Transiting Exoplanet Survey Satellite} (\tess) \citep[e.g.,][]{Muirhead:2018} missions. As a result, detailed and precise knowledge of M-dwarf chemical compositions has become key to constraining the properties, formation scenarios, and atmospheric conditions of potentially habitable exoplanets observable with the \textsl{James Webb Space Telescope} ({\sl JWST;} \citealt{Clampin2008}). 

Advances in instrumentation and the implementation of several spectroscopic surveys in the past decade, such as the Sloan Digital Sky Survey \citep[SDSS;][]{Eisenstein:2011,Blanton:2017} and the Large Sky Area Multi-Object Fiber Spectroscopic Telescope \citep[LAMOST;][]{Zhao:2012}, have dramatically increased the sample of known M dwarfs \citep{West:2011,Guo:2015} with spectroscopic catalogs of over 70,000 sources, enabling studies of M-dwarf abundances on a Galactic scale.
The Apache Point Observatory Galactic Evolution Experiment (APOGEE; \citealt{Majewski:2015}) survey, as part of the SDSS III/IV mission, has introduced the largest sample of M dwarfs observed with high-resolution spectroscopy \citep{Desphande:2013}. APOGEE pipeline measurements of $\teff$ and $\feh$ \citep{Perez:2016} for M dwarfs have been determined to precisions of 100\,K and 0.18\,dex down to $T_{\rm eff}\sim3550\,$K using atmosphere models \citep{Schmidt:2016}. 

Elemental abundance measurements from high-resolution
spectra of F, G, and K stars have achieved extremely high
precision (down to $0.01-0.03$\,dex; \citealt{Nissen2018}) 
enabled by improvements in atmosphere models including
realistic assumptions of 3D local thermodynamic equilibrium 
\citep{Asplund2005}, and differential abundance techniques using 
equivalent widths \citep{Bedell2014}. However, the determination 
of precise metallicities for M dwarfs has remained a long-standing challenge due to the formation of diatomic and triatomic molecules at M-dwarf temperatures, with absorption from TiO and VO in the optical, H$_2$O and CO in the infrared, and hydrides (FeH, CaH, CrH, MgH, etc.) present in the spectra of the latest spectral types \citep{Allard1997}.
Atmospheric models often fail to reproduce these spectral features \citep[e.g.,][]{Mann:2013c} because of incomplete line lists and opacities. The presence of millions of weak, blended transitions, and the absence of a clear continuum, contribute to making it difficult to deconvolve individual features and extract line strengths from equivalent widths. The combination of these effects limits our ability to explore the information content of high-resolution spectra using traditional methods.

A number of studies focused on improving precisions of
M-dwarf metallicity have used systems of M dwarfs in
common proper motion with an FGK star and strong, isolated
lines in the spectra of the M dwarf \citep[e.g.,][]{Terrien:2012,RojasAyala:2010,Newton:2014,Neves:2014,Lindgren:2016} to develop precise empirical relations (as good as $\sim$0.07\,dex). 
However these metallicity calibrations do not take advantage of the full wavelength coverage available, nor information about the overall spectral shape often used to determine $\teff$ and spectral type. Furthermore, earlier calibrations are generally based on moderate-resolution data \citep[with some exceptions:][]{Neves:2014, Lindgren:2016} that fail to utilize the greater spectral information provided by APOGEE's resolution.

In this work we build a data-driven model for M-dwarf APOGEE spectra with \thecannon\ \citep{Ness:2015,Casey:2016,Ho:2017a,Behmard2019}--a fully empirical model that employs no line lists or radiative transfer models. 
\thecannon\ is a \emph{generative model} that parameterizes the flux at each pixel of a spectrum in terms of a set of stellar labels (a flexible number of parameters chosen by the user; described in more detail in Section \ref{sec:methods}). The model in this sense is used to \emph{transfer} labels from spectra for which we know parameters to those for which we do not. This data-driven approach effectively circumvents the challenges of physically modeling the atmosphere of a star (and common issues associated such as incomplete line lists or opacities), provided that we have a subset of spectra in the dataset with known (and very accurately measured) \emph{reference labels} possibly measured from other data. 

The data-driven approach of \thecannon\ is ideal in certain cases:
if stellar labels are known for a small number of stars
but there are spectra taken for many more;
if it is computationally expensive to obtain labels for a star, and
there are many stars that need labels;
or if there are spectral models or techniques that work in one wavelength range or resolution but not in another.
Existing methods to model M-dwarf spectra in the
near infrared at high resolution are computationally expensive,
and often calibrated over a narrow range of $\teff$ and/or metallicity. 
\thecannon\ thus fills this niche: it does not require
the use of specific lines or opacity information that may be 
missing from the models; instead it allows us to determine
labels from a lot of low-level metallicity information present in
thousands of lines, and as we demonstrate, it does so with very
good precision.

Here we take M-dwarf labels from samples of well-characterized stars that are present in the
SDSS-IV APOGEE sample, and use those labels to train a model and label all of the M dwarfs observed by
SDSS-IV APOGEE. One set of labels are physical parameters (effective temperatures and metallicities), the other set of labels are spectral types. 

This paper is organized as follows: in Section \ref{sec:data} we describe the technical specifications of the data from the APOGEE and \gaia\ surveys, as well as previous studies of M dwarfs in APOGEE.
Section \ref{sec:methods} describes our model implementation using \thecannon\ framework, and Section \ref{sec:sample_selection} describes our sample selection and derivation of training parameters.
In Section \ref{sec:results} we present our experimental results, evaluate the predictive accuracy of our models, apply our model to a selected test sample of nearly 6000 sources, and examine the validity of our parameters against color-temperature relations and metallicities of binary pairs.
Finally, in Section \ref{sec:discussion} we discuss model performance, future improvements, and implications of our results.

\section{Data} \label{sec:data}

The APOGEE survey is a high-resolution ($R\sim22,500$), $H$-band ($1.5-1.7\mu$m), multi-epoch survey that has observed over 250,000 stellar spectra up to its fourteenth data release (DR14; \citealt{Abolfathi:2017}). Fundamental parameters for each of these stars are estimated by the APOGEE Stellar Parameter and Chemical Abundances Pipeline (ASPCAP; \citealt{Perez:2016}), which employs a $\chi^2$ fitting procedure using the FERRE code to fit radiative transfer models and determine atmospheric parameters, 15 chemical abundances, and microturbulence parameters \citep{Meszaros:2012}. 
The pipeline uses MARCS plane-parallel/spherical models \citep{Gustafsson:2008} for
low temperatures ($2800<\teff<3500$\,K), and ATLAS9
plane-parallel models \citep{Castelli:2004} for higher
temperatures ($\teff\geq3500$\,K).

APOGEE is primarily designed to target bright stellar populations, particularly red giants, with dereddened photometry and color cutoffs of $7 \leq H \leq 13.8$ and $[J-K]_0 \geq 0.5$ \citep{Zasowski:2013}, with the objective of studying Galactic composition and evolution. However, numerous cool, main-sequence sources have also been observed either as targets proposed by the APOGEE M-dwarf ancilliary survey ($\sim$1,200 sources; \citealt{Desphande:2013}), or serendipitously. 

A number of studies out of the M-dwarf ancillary survey
have already been conducted to measure reliable fundamental
atmospheric parameters and make kinematic measurements
using spectral synthesis of atmospheric model grids. 
These studies include \citet{Desphande:2013} and \citet{Gilhool:2018} which have studied the radial and rotational kinematics for 700+ sources;
\citet{Souto:2017} and \citet{Souto:2018} which have modeled three exoplanet-hosting M dwarfs (Kepler-138, Kepler-168, and Ross-128), determining $\teff$/$\logg$/metallicity + 13 elemental abundances;
\citet{Rajpurohit:2018} which tested BT-Settl \citet{Allard:2012} and MARCS \citet{Gustafsson:2008} model grids on 45 M dwarfs to estimate $\teff$/$\logg$/metallicity;
and \citet{Skinner:2018} which identified and measured mass ratios and radial velocities for 44 M-dwarf spectroscopic binaries.
This work complements existing studies by producing a model-independent catalog of
spectroscopic temperatures and metallicities to test against
model predictions for the entire APOGEE M-dwarf sample,
which we quantify to contain at least 10,000 sources to
date (DR14).

The ASPCAP pipeline releases several types of data files, with various levels of processing: {\tt\string ap1D} (the raw one-dimensional spectra for individual visits), {\tt\string apVisit} (the individual visit spectra with telluric subtraction), {\tt\string apStar} (the co-added {\tt\string apVisit} spectra), and {\tt\string aspcapStar}, which contains the pseudo-continuum-normalized, rest-frame-shifted, co-added spectrum of all observed epochs (see \citealt{Perez:2016} for a complete description of the pipeline). We use the last dataset for our study.
In previous work it has been recommended to use an alternative pseudo-continuum normalization \citep{Ness:2015}, but we did not find obvious issues with the normalization in our analysis, so we retain the survey pipeline outputs.

\section{Method} \label{sec:methods}

\thecannon\ is a regression model that relies on two assumptions: first, that sources with identical labels have near-identical flux at each wavelength pixel; and second, that the expected flux at each pixel varies continuously with change in label. 

Inferring the label of a star with such a model requires two steps: first, the \emph{training step} in which a generative model describing the probability density function of the flux is constructed at each pixel from the set of spectra with known reference labels; and second, the \emph{test step} in which the model is applied to determine the labels of a spectrum.

Following the procedure of \citet{Ness:2015} and \citet{Ho:2017a} we adopt a simple linear model that assumes that the flux at each pixel of the spectrum can be parameterized as a function of a label vector $\ell$ and coefficient vector \emph{$\theta$}. For each star \emph{n}, at wavelength pixel \emph{$\lambda$}, we assume that the measured flux for a star at a given pixel is the sum of the coefficient and label product, and observational noise:
\begin{equation}
	f_{n\lambda} = \theta_{\lambda}^{T} \cdot \ell_{n} + N_{\lambda} \label{eqn:cannon}
\end{equation} 
Here we use the noise model $N_{\lambda}=[s_{\lambda}^2 + \sigma_{n\lambda}^2]\xi_{n\lambda}$, where the bracketed term is the root mean sum of the intrinsic scatter of the model at each pixel \emph{$s_{\lambda}$}, and the uncertainty due to instrumental effects \emph{$\sigma_{n\lambda}$}, which is then multiplied by a Gaussian random number $\xi_{n\lambda} \sim \mathcal{N} (0,1)$. Equation (\ref{eqn:cannon}) corresponds to the single-pixel log-likelihood function
\begin{equation}
	\ln p(f_{n\lambda}|\theta^T_{\lambda}, \ell_n, s^2_\lambda) = 
	-\frac{1}{2}\frac{[f_{n\lambda} - \theta_{\lambda}^{T} \cdot \ell_{n}]^2}{s_{\lambda}^2 + \sigma_{n\lambda}^2} - \frac{1}{2} \ln(s_{\lambda}^2 + \sigma_{n\lambda}^2) 
	\label{eqn:likelihood}
\end{equation}
which gives the probability density function of the measured flux, given the labels, coefficients, and scatters.

We apply a quadratic parameterization of the model such that the label vectors for the two models are all combinations of reference labels up to second order:
\begin{eqnarray}
	\ell_{n} &=& [1, \, SPT, \, SPT^{2}] \label{eqn:spt}
	\\
	\ell_{n} &=& [1, \, \teff, \, \feh, \, \teff^2, \, \teff \cdot \feh, \, \feh^{2}] \label{eqn:physical}
	.
\end{eqnarray}
Equation (\ref{eqn:spt}) is the label vector for the spectral type model, and equation (\ref{eqn:physical}) is the label vector for the physical parameter model; the first element ``1'' is included to allow flexibility for a linear offset to the model. We find that a second-order parameterization is sufficient for reproducing the flux of each spectrum to 1\% accuracy, as discussed further in Section \ref{subsec:mann_results}.

The \emph{training step} consists of optimizing the likelihood function (Equation \ref{eqn:likelihood}) for the coefficient vector and scatter ($\theta_{\lambda}$ and $s_{\lambda}$) given the fixed label vector (\emph{$\ell_n$}) constructed from the reference labels. The \emph{test step} consists of optimizing the likelihood function for the labels at fixed $\theta_{\lambda}$ and $s_{\lambda}$ obtained in the training step (see \citealt{Ness:2015} for further description). 
In the training step, the regression is designed to predict spectral pixels
given labels, by learning zeroth, first, and second derivatives of the data with respect to
the labels. In the test step, the regression is designed to predict labels given the spectral
derivatives.

\section{Sample Selection} \label{sec:sample_selection}

\thecannon\ model can in principle be trained on any physical or empirical labels available beyond those that typically parameterize theoretical atmospheric models ($\teff$, log$g$, [M/H], etc.), such as additional physical parameters (e.g., mass/age \citealt{Ho:2017b}) or empirical proxies for physical parameters (e.g., spectral types, colors, magnitudes), giving a wide range of flexibility to the model. 
However, choosing a training sample with high-quality labels is critical to its performance.
Limitations of \thecannon\ include that test (output) labels are only accurate if the training labels are accurate, and only precise if the training labels are measured consistently across the training sample.
It is also critical to have a training sample with the dynamic range to span the entire parameter space of interest, as \thecannon\ does not extrapolate well outside the parameter space of the training sample.
Finally, \thecannon\ assumes that the dependence of the spectrum on labels is continuous and smooth---and in this implementation is, well approximated by quadratic functions. If that is not true, there will be features that \thecannon\ cannot reproduce.

For the purpose of this study, we have constructed two different training samples: first a one-dimensional \emph{spectral type model}, and second, a two-dimensional \emph{physical parameter model}, which describes the temperature and metallicity. The choice of training labels, dimensionality of our data set, and requirements for a good training set are discussed further in Section \ref{sec:discussion}.

\subsection{Spectral Type Training Sample}

\begin{figure*}
	\begin{center}
	\includegraphics[width=12cm]{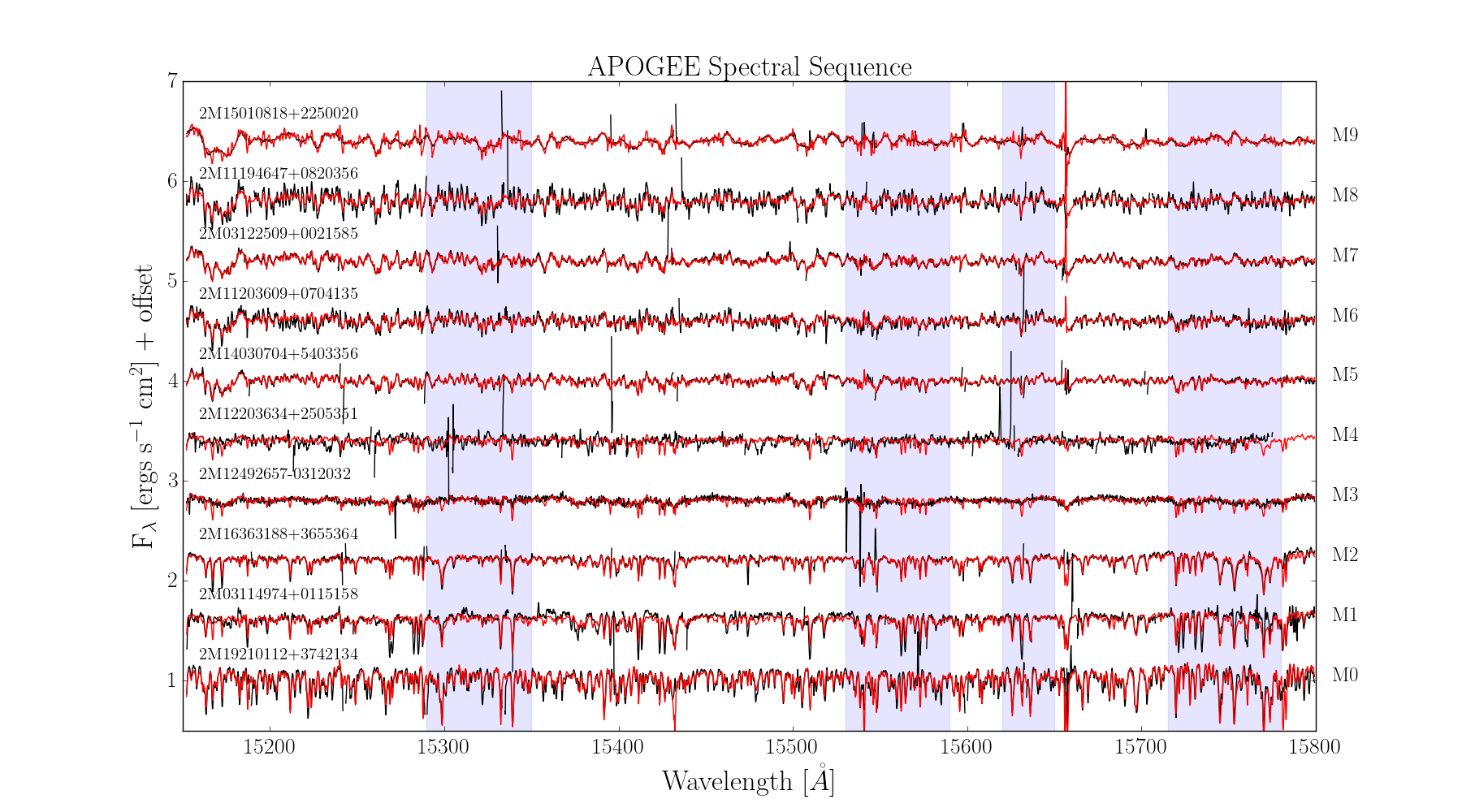}
	\includegraphics[width=12cm]{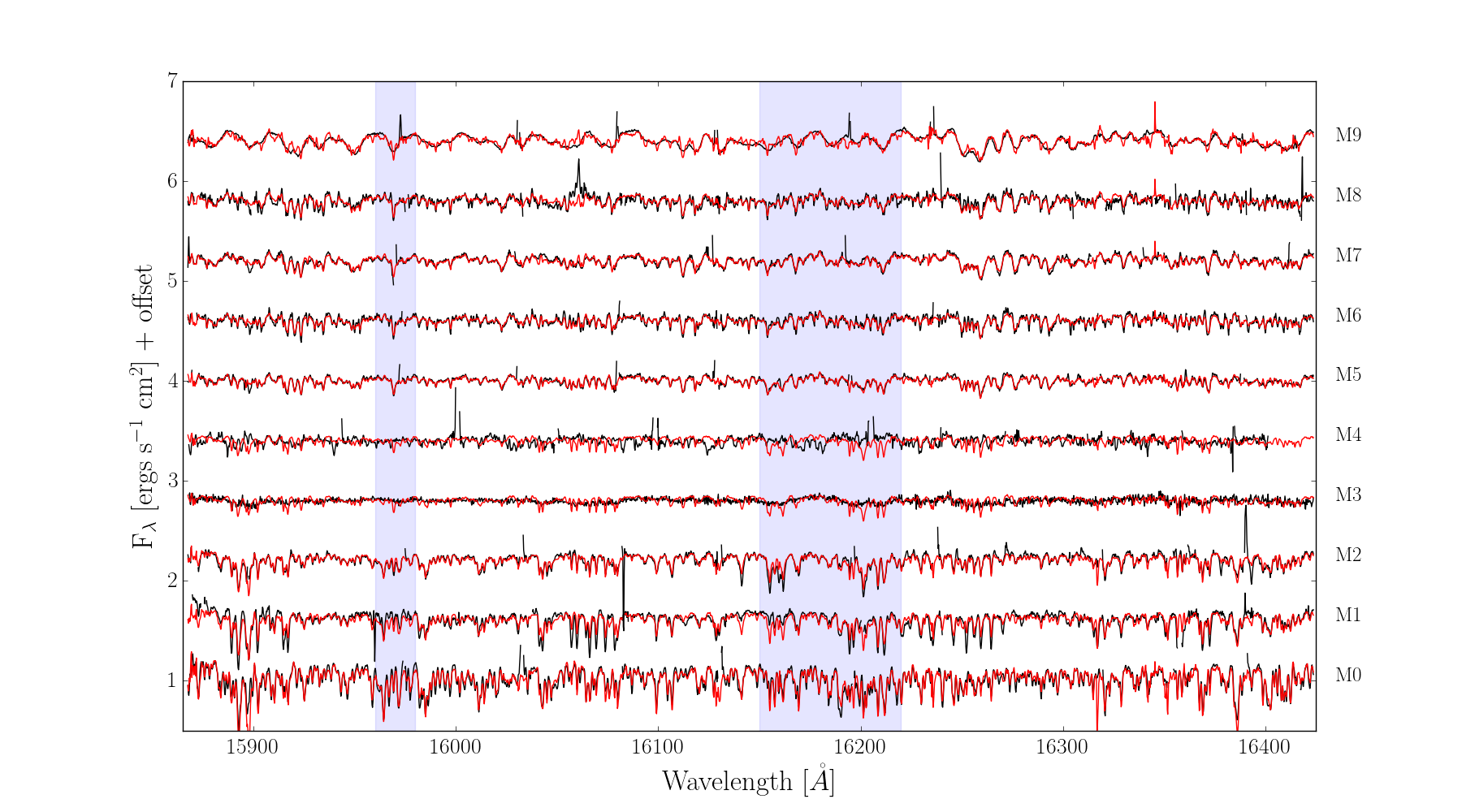}
	\includegraphics[width=12cm]{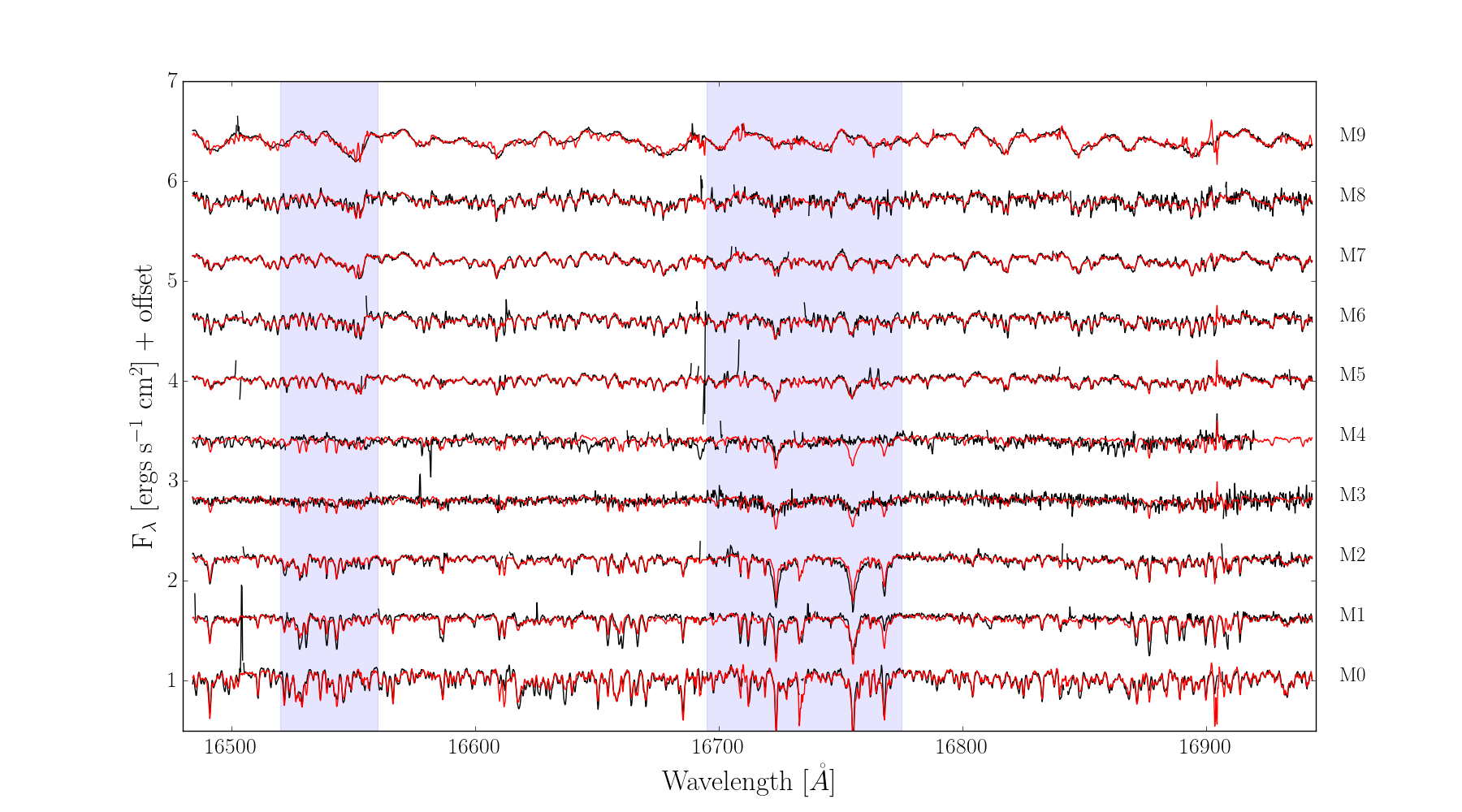}
	\end{center}
	\caption{Spectral sequence of dwarfs in training set M0-M9; separate plots show three detector chips of APOGEE spectrum (black), the best fit Cannon trained model (red) with highlighted spectral type sensitive regions identified in \citealt{Desphande:2013}.} \label{fig:sp_sequence}
\end{figure*}

The spectral type training sample consists of 51 sources, spanning M0$-$M9 cross-matched from the catalog of \citet{West:2011} (hereafter W11) that contains 78,841 M dwarfs from SDSS. For each source in the catalog, spectral types were determined both through an automated routine for comparing spectral type templates to data using \textsl{The Hammer} \citep{Covey:2007} and by visual inspection to a reported accuracy of $\pm$1 type. A spectral sequence of spectra from the training sample
spanning M0-M9 is shown in Figure \ref{fig:sp_sequence}.

\subsection{Physical Parameter Training Sample}

The physical parameter training sample consists of 87 sources with reference labels distributed over $2859 <T_{\rm eff} < 4131$K, and $-0.48 < \feh < 0.49$\,dex, 41 sources of which are drawn from \citet{Mann:2015} (hereafter M15), and 46 of which are part of a previously unpublished extension sample to M15, analyzed using similar data and identical techniques to M15. The major difference in the extension sample is that the sample had lower-quality or no parallaxes (prior to \gaia\ data) and hence were omitted from the M15 study and were less vetted for binarity than the M15 sample (however, all sources in the training sample were visually inspected by color-magnitude position for binarity before addition).

The M15 catalog in total contains 183 sources and the extension sample another 500 stars. Both samples were primarily selected from the proper-motion-selected CONCH-SHELL \citep{Gaidos:2013} M-dwarf catalog. All targets have low-resolution optical spectra from the SNIFS spectrograph \citep{Lantz:2004} and infrared spectra taken with the SpeX Spectrograph \citep{Rayner:2003}, which have been combined to estimate largely empirical bolometric fluxes. Effective temperatures have been estimated by comparing the SNIFS spectra to BT-SETTL atmospheric models \citep{Allard:2011}. A subsample of 29 sources with measured angular diameters from long-baseline optical interferometry \citep{Boyajian:2012} are used to calibrate the model comparison, including masking of spectral regions poorly reproduced by the model spectra \citep{Mann:2013c}. Based on the difference between assigned $\teff$ values and those from angular diameters, absolute uncertainty on $\teff$ is estimated to be 60K in $\teff$, although the relative uncertainty is likely a factor of $\simeq$2 better.

Iron abundances ($\feh$) are assigned to the physical parameter sample based on the strength of metal-sensitive lines in the near-infrared SpeX spectra \citep{RojasAyala:2010} using the calibration from \citet{Mann:2013a}. The relation between these lines and an absolute $\feh$ scale is calibrated using wide binaries containing an F, G, or K-type primary and an M-dwarf companion, under the assumption that binaries formed from the same molecular cloud and therefore have the same metallicity \citep{Bonfils:2005}. Uncertainties are estimated to be $\simeq$0.08\,dex based on irreducible scatter in the empirical relation between selected lines and the assigned $\feh$ from the primary star. As with $\teff$, relative errors on $\feh$ are smaller, estimated to be 0.04--0.06\,dex over most of the temperature and metallicity range considered here.

We note that surface gravity is not included as a training
label. The reason for this is that for main-sequence M dwarfs,
the parameter is almost entirely redundant with metallicity.
The properties of M dwarfs, unlike those of their more massive
counterparts, do not change measurably over the age of the
universe after arriving at the zero-age main sequence.
Hence perfect knowledge of abundances and $\teff$ for an M dwarf should uniquely determine its surface gravity, position on a color-magnitude diagram, and overall luminosity. While we only had $\feh$ for the training sample, for the uncertainties considered here, lack of information about [$\alpha$/Fe] or specific abundances will only be important compared to other uncertainties in extreme cases (e.g., carbon stars).

\section{Experiments and Results} \label{sec:results}

\subsection{Temperature/Metallicity Model \label{subsec:mann_results}}

\begin{figure}
	\includegraphics[width=\linewidth]{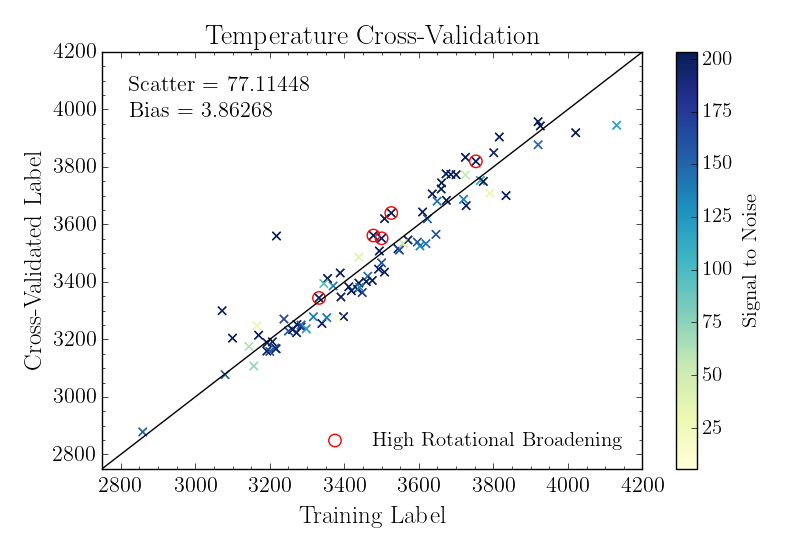}
	\includegraphics[width=\linewidth]{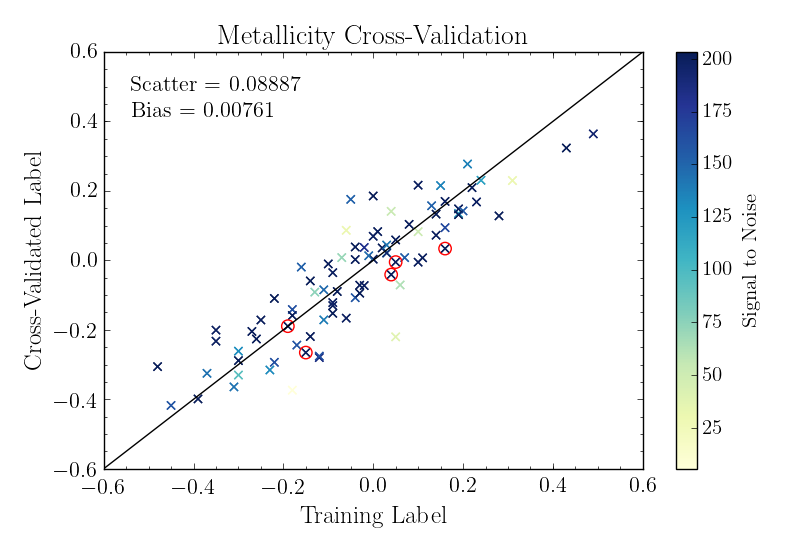}
	\caption{Consistency test for the Mann-trained temperature/metallicity type model; self-test (top), and leave-one-out cross-validation (bottom). Sources with strong rotational broadening (2M08155393+3136392, 2M09174473+4612246, 2M10331367+3409120, 2M15594729+4403595, 2M13400879+4346380), as identified by visual inspection, are marked with red circles. Predictive accuracy, as computed from the scatter in cross-validation is 77K in temperature and 0.09 dex in metallicity.} 
	\label{fig:mann_validation}
\end{figure}

\begin{deluxetable*}{cccccccc}
\tablecaption{Metallicities of APOGEE M dwarfs in wide binaries. Reported in this table are the ASPCAP [M/H] and \cannon\ [Fe/H] of the M dwarfs, as well as the metallicity values of their binary companions. The source/survey of the metallicity and spectral classes of the companions are also given. \label{table:binaries}}
\tabletypesize{\scriptsize} 
\tablehead{\colhead{Mdwarf Gaia ID}      & \colhead{Companion Gaia ID}   & \colhead{Mdwarf [M/H]} & \colhead{Mdwarf [Fe/H]} & \colhead{Companion [Fe/H]} & \colhead{Companion source} & \colhead{Separation (AU)}}
	\startdata
	1476768433933264128 & 1476762627137479552 & -0.18      & 0.01        & 0.10           & APOGEE Cannon M   & 16091      \\
	1476646285062821888 & 1476646250703083264 & -0.02      & 0.18        & 0.27           & APOGEE Cannon M   & 4155       \\
	4008743154907642496 & 4008743086188165248 & -0.20      & 0.02        & 0.09           & APOGEE Cannon M   & 11855      \\
	3053911383155592832 & 3053907221325174144 & -0.25      & -0.15       & -0.15          & APOGEE Cannon M   & 12884      \\
	1475867727751525632 & 1475867727751525504 & -0.25      & -0.15       & -0.13          & APOGEE Cannon M   & 5091       \\
	933708229245319552  & 933708164821511424  & -0.23      & 0.11        & 0.01           & APOGEE Cannon M   & 2245       \\
	3888520938618496768 & 3888520938618115456 & -0.16      & -0.15       & -0.14          & APOGEE Cannon M   & 1128       \\
	3888309828090932608 & 3888309832385988992 & -0.24      & -0.01       & 0.04           & APOGEE Cannon M   & 2499       \\
	374137560388179328  & 374137560388178304  & -0.34      & -0.24       & -0.34          & LAMOST FGK        & 5888       \\
	4031346880591589248 & 4031346914951327104 & -0.33      & -0.25       & -0.18          & LAMOST FGK        & 6637       \\
	3443427533502334592 & 3443427533502335744 & 0.06       & 0.20        & 0.32           & LAMOST FGK        & 3151       \\
	1605991802162861696 & 1605991767803123200 & -0.16      & -0.03       & -0.06          & LAMOST FGK        & 1043       \\
	4418850890305439872 & 4418851263967008000 & -0.18      & 0.11        & 0.27           & APOGEE ASPCAP FGK & 4920       \\
	1381455722290898176 & 1381408782592655616 & -0.13      & -0.02       & -0.08          & APOGEE ASPCAP FGK & 2401       \\
	2687603208839189376 & 2687603311918406528 & -0.12      & 0.00        & -0.03          & APOGEE ASPCAP FGK & 11201    
	\enddata
\end{deluxetable*}

\begin{deluxetable*}{ccc|ccc|ccc|c}
\tablecaption{\cannon\ results for Temperature/Metallicity model; see online journal for full table. Reported uncertainties for the M15 training sample labels are $\pm$60\,K/0.08\,dex, and reported uncertainties for \thecannon\ model based on the cross-validation scatter are $\pm$77\,K/0.09\,dex. \label{table:mann_results}}
\tabletypesize{\scriptsize}
\tablehead{
\multicolumn{3}{c}{\underline{Designation}} & \multicolumn{3}{c}{\underline{Temperature (K)}} & \multicolumn{3}{c}{\underline{Metallicity (dex)}} & \multicolumn{1}{c}{\underline{Model Fit}} \\
	  \colhead{2MASS ID} & \colhead{RA}   & \colhead{DEC}
    & \colhead{Training} & \colhead{Test} & \colhead{LOOCV}
    & \colhead{Training} & \colhead{Test} & \colhead{LOOCV} 
    & \colhead{Test $\chi^2$} 
} 
	\startdata
	2M00182256+4401222 & 4.59542   & 44.02278  & 3603        & 3538       & 3525        & -0.3       & -0.28     & -0.26      & 15304   \\
	2M00182549+4401376 & 4.60779   & 44.02734  & 3218        & 3528       & 3560        & -0.3       & -0.29     & -0.29      & 21273   \\
	2M00285391+5022330 & 7.22488   & 50.37588  & 3207        & 3190       & 3192        & 0.11       & 0.04      & 0.01       & 31790   \\
	2M00401001+0308050 & 10.04169  & 3.13473   & 3725        & 3777       & 3772        & 0.04       & 0.12      & 0.14       & 9830    \\
	2M00580115+3919111 & 14.50482  & 39.31977  & 3157        & 3100       & 3107        & -0.07      & -0.02     & 0.01       & 13337   \\
	2M01232542+1638384 & 20.8559   & 16.64401  & 3272        & 3225       & 3223        & 0.1        & 0.02      & -0.01      & 34797   \\
	2M02001278+1303112 & 30.05402  & 13.05196  & 3080        & 3059       & 3077        & -0.16      & -0.1      & -0.02      & 33210   \\
	2M02361535+0652191 & 39.06358  & 6.87167   & 3284        & 3241       & 3243        & -0.12      & -0.28     & -0.28      & 48445   \\
	2M03044335+6144097 & 46.18104  & 61.73583  & 3500        & 3466       & 3466        & -0.12      & -0.26     & -0.28      & 40739   \\
	2M03553688+5214291 & 58.90373  & 52.2414   & 3435        & 3386       & 3375        & -0.35      & -0.28     & -0.2       & 53870   \\
	2M04125880+5236421 & 63.24499  & 52.61165  & 3100        & 3183       & 3204        & -0.04      & -0.01     & 0.0        & 46007   \\
	2M04310001+3647548 & 67.75003  & 36.79855  & 3419        & 3371       & 3369        & 0.08       & 0.11      & 0.1        & 20602   \\
	2M05222053+3031097 & 80.58561  & 30.51933  & 3389        & 3423       & 3431        & 0.28       & 0.18      & 0.13       & 19072   \\
	2M05312734-0340356 & 82.86417  & -3.67722  & 3801        & 3814       & 3849        & 0.49       & 0.5       & 0.36       & 28209   \\
	2M05413073+5329239 & 85.37804  & 53.48987  & 3765        & 3751       & 3753        & 0.19       & 0.15      & 0.14       & 17842   \\
	2M05420897+1229252 & 85.53833  & 12.48956  & 3250        & 3220       & 3229        & -0.22      & -0.33     & -0.29      & 33868   \\
	2M06000351+0242236 & 90.01458  & 2.70657   & 3214        & 3170       & 3170        & 0.07       & 0.03      & 0.01       & 21044   \\
	2M06011106+5935508 & 90.2961   & 59.59713  & 3340        & 3259       & 3255        & -0.09      & -0.05     & -0.04      & 23329   \\
	2M06112610+1032599 & 92.8588   & 10.54998  & 3636        & 3720       & 3706        & -0.39      & -0.43     & -0.4       & 36898   \\
	2M06544902+3316058 & 103.70411 & 33.26823  & 3448        & 3368       & 3363        & -0.02      & 0.03      & 0.04       & 19412   \\
	2M07171706-0501031 & 109.32108 & -5.01754  & 3193        & 3175       & 3188        & -0.09      & -0.15     & -0.13      & 103000  \\
	2M07272450+0513329 & 111.852   & 5.2263    & 3317        & 3279       & 3279        & -0.11      & -0.18     & -0.17      & 23615   \\
	2M07444018+0333089 & 116.16744 & 3.55252   & 3217        & 3174       & 3167        & 0.23       & 0.25      & 0.17       & 34181   \\
	2M08031949+5250387 & 120.83131 & 52.84402  & 3508        & 3617       & 3620        & -0.26      & -0.23     & -0.23      & 17491   \\
	\nodata & \nodata & \nodata & \nodata & \nodata & \nodata & \nodata & \nodata & \nodata & \nodata
	\enddata
\end{deluxetable*}

\begin{figure*}
	\includegraphics[width=\linewidth]{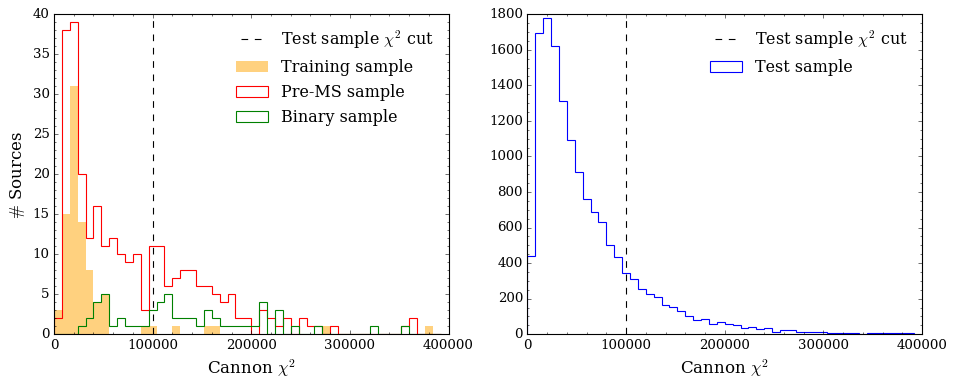}
	\caption{\textit{Left:} the distribution of fits for the training sample and known samples of pre-main-sequence stars (\citealt{Cottaar:2014}) and binary sources \citep{ElBadry:2018,Skinner:2018}. \textit{Right:} the distribution of $\chi^2$ fits for all 14,828 sources in the APOGEE--\gaia\ cross-match, with color cuts $1<G_{BP}-G_{RP}<6$ and $7.5<M_{G}<20$ and $\varpi>0$. We apply a quality cut of $\chi^2 < 100,000$ to the test sample for those sources we report as ``safe''.}
	\label{fig:chi_dist}
\end{figure*}

For the physical parameter model, we trained \thecannon\ on 87 M dwarfs with two-dimensional temperature/metallicity labels, to a precision of 77K/0.09dex as estimated by the cross-validation scatter, similar to the uncertainties on the original training sample of 60 K/0.08 dex.
We note for this model that 5 out of 87 sources show possible rotational line broadening identified by visual inspection (as indicated by the red circles in Figure~\ref{fig:mann_validation}), while the remaining sources show no obvious broadening. We note that these broadened sources have high $\chi^2$ values (those sources with $\chi^2 > 80,000$ in Figure \ref{fig:chi_dist}), and that the labels for these five sources are biased by an average of $+65\,$K and $-0.08\,$dex. However, removing them from the training sample does not significantly change the overall scatter and bias of the model. For the model overall, the cross-validation bias is +4K/+0.008 dex with the rapid rotators included in the training set, and the bias is +5K/+0.01 dex when they are excluded. Hence we do not remove them from the training sample.

To assess the validity of our model's labels we used a leave-one-out cross-validation (LOOCV) test, in which we train a model on all sources but \emph{n}, then apply the \emph{N-1} source-trained model to obtain the labels for star \emph{n}. Precision (scatter) and bias of the model for each test are calculated as the standard deviation and mean of the difference in training and test (or LOOCV) labels respectively (Figure \ref{fig:mann_validation}). Since the LOOCV test evaluates how well the model reproduces the training values \emph{and} penalizes the model for overfitting, we adopt the LOOCV scatter as the estimate of the model's precision. The set of training, test, and cross-validated labels for each training source is reported in Table \ref{table:mann_results}.

Another mode of analysis we can utilize with \thecannon\ is how the derivative of the model changes with respect to given training parameters, which makes our model interpretable for discovering or verifying atomic or molecular lines with strong dependence on different physical parameters. The top two panels of Figures \ref{fig:demo_teff} and \ref{fig:demo_feh} show two example spectra and model fits for two different temperatures (Figure \ref{fig:demo_teff}), and two different metallicities (Figure \ref{fig:demo_feh}), with atomic and molecular features identified by the abundance analysis of \citet{Souto:2017}. The bottom panels of Figures \ref{fig:demo_teff} and \ref{fig:demo_feh} show the derivative of flux with respect to temperature and metallicity at each pixel, taken at the median training values. In order to evaluate which spectral features show statistically significant change with respect to input label, we compute the error of the derivative at each pixel using a jackknife statistic (with a 1-$\sigma$ level overplotted in red):
\begin{equation}
	\sigma_{\theta,m}^2 = \frac{N-1}{N} \sum^N_{n=1} (\theta_{/n}^m - \theta^m)^2 
\end{equation}
where $\sigma_{\theta,m}$ is the error at pixel \emph{m}, \emph{N} is the total number of stars in the sample indexed by \emph{n}, $\theta$ is the coefficient vector trained on all \emph{N} sources, and $\theta_{/n}$ is the coefficient vector trained on \emph{N-1} sources excluding star \emph{n}. A summary of identified lines with derivative values greater than $2\sigma_{\rm jackknife}$ is given in Table \ref{table:teff_derivative} and Table \ref{table:feh_derivative}.

\FloatBarrier
\startlongtable
\begin{deluxetable}{ccccc}{H}
\tablecaption{Strong $\teff$-correlated atomic and molecular lines for all features with significance greater than $2\sigma_{\rm jackknife}$. `Significance' quantifies the number of standard deviations that the derivative of the line is significant to (with negative corresponding to negative derivatives).}
\label{table:teff_derivative}
\tabletypesize{\scriptsize}
\tablehead{\colhead{Species} & \colhead{Wavelength} & \colhead{Derivative} & \colhead{$\sigma_{\rm jackknife}$} & \colhead{Significance}}
	\startdata
	FeH & 16107.127 & -0.002 & 0.001 & -2.78 \\
	FeH & 16245.463 & -0.005 & 0.002 & -3.214 \\
	FeH & 16271.519 & -0.002 & 0.001 & -2.013 \\
	FeH & 16284.562 & -0.006 & 0.001 & -5.807 \\
	FeH & 16377.066 & -0.005 & 0.001 & -4.554 \\
	FeH & 16741.489 & -0.002 & 0.001 & -2.873 \\
	FeH & 16812.415 & -0.004 & 0.001 & -4.241 \\
	FeH & 16813.808 & -0.006 & 0.001 & -3.954 \\
	FeH & 16922.405 & -0.003 & 0.001 & -3.867 \\
	FeH & 16934.801 & -0.004 & 0.001 & -3.591 \\
	OH & 15278.267 & 0.006 & 0.003 & 2.47 \\
	OH & 15280.8 & 0.004 & 0.002 & 2.007 \\
	OH & 15391.186 & 0.006 & 0.003 & 2.127 \\
	OH & 15407.355 & 0.01 & 0.002 & 4.836 \\
	OH & 15409.271 & 0.008 & 0.002 & 4.683 \\
	OH & 15505.797 & 0.009 & 0.003 & 3.511 \\
	OH & 15560.305 & 0.008 & 0.002 & 3.865 \\
	OH & 15565.895 & 0.005 & 0.002 & 2.505 \\
	OH & 15568.691 & 0.007 & 0.001 & 4.929 \\
	OH & 15572.133 & 0.008 & 0.002 & 4.393 \\
	OH & 16052.7 & 0.008 & 0.002 & 3.517 \\
	OH & 16055.362 & 0.006 & 0.002 & 3.7 \\
	OH & 16061.796 & 0.007 & 0.003 & 2.502 \\
	OH & 16065.124 & 0.009 & 0.002 & 4.214 \\
	OH & 16069.564 & 0.008 & 0.003 & 3.275 \\
	OH & 16074.227 & 0.007 & 0.002 & 3.338 \\
	OH & 16190.345 & 0.007 & 0.003 & 2.498 \\
	OH & 16192.134 & 0.008 & 0.002 & 4.012 \\
	OH & 16203.995 & 0.006 & 0.002 & 3.039 \\
	OH & 16207.129 & 0.004 & 0.002 & 2.196 \\
	OH & 16352.196 & 0.007 & 0.002 & 3.609 \\
	OH & 16354.682 & 0.011 & 0.003 & 4.167 \\
	OH & 16364.626 & 0.007 & 0.003 & 2.661 \\
	OH & 16368.244 & 0.008 & 0.002 & 4.303 \\
	OH & 16581.281 & 0.008 & 0.002 & 4.41 \\
	OH & 16581.74 & 0.004 & 0.001 & 4.842 \\
	OH & 16866.621 & 0.005 & 0.001 & 3.169 \\
	OH & 16871.982 & 0.012 & 0.002 & 4.892 \\
	OH & 16878.976 & 0.006 & 0.002 & 3.807 \\
	OH & 16884.573 & 0.005 & 0.002 & 3.066 \\
	OH & 16886.206 & 0.006 & 0.001 & 5.131 \\
	OH & 16895.074 & 0.005 & 0.002 & 2.197 \\
	Fe I & 15207.509 & -0.007 & 0.003 & -2.622 \\
	Fe I & 15648.478 & -0.006 & 0.002 & -3.273 \\
	Fe I & 15692.643 & -0.008 & 0.002 & -4.056 \\
	Si I & 15960.044 & -0.009 & 0.004 & -2.523 \\
	Si I & 16094.893 & -0.014 & 0.003 & -4.368 \\
	Si I & 16680.77 & -0.01 & 0.003 & -3.475 \\
	K I & 15162.824 & -0.006 & 0.002 & -2.294 \\
	Ti I & 15715.641 & 0.006 & 0.002 & 2.894 \\
	Ti I & 16634.973 & 0.003 & 0.001 & 4.224 \\
	V I & 15923.924 & 0.003 & 0.001 & 3.6 \\
	Mn I & 15159.054 & -0.006 & 0.002 & -3.123 \\
	\enddata
\end{deluxetable} 
\FloatBarrier

\FloatBarrier
\begin{deluxetable}{ccccc}{H}
\tablecaption{Strong $\feh$-correlated atomic and molecular lines for all features with significance greater than $2\sigma_{\rm jackknife}$.}
\tabletypesize{\scriptsize}
\tablehead{\colhead{Species} & \colhead{Wavelength} & \colhead{Derivative} & \colhead{$\sigma_{\rm jackknife}$} & \colhead{Significance}}
	\startdata
	FeH & 16114.027 & 0.004 & 0.001 & 2.673 \\
	FeH & 16574.64 & 0.003 & 0.001 & 2.356 \\
	FeH & 16694.372 & 0.003 & 0.001 & 2.193 \\
	FeH & 16814.041 & 0.004 & 0.002 & 2.513 \\
	OH & 16052.7 & 0.005 & 0.002 & 2.034 \\
	OH & 16055.362 & 0.006 & 0.002 & 2.805 \\
	OH & 16065.124 & 0.007 & 0.003 & 2.364 \\
	OH & 16871.982 & 0.007 & 0.003 & 2.306 \\
	OH & 16886.206 & 0.004 & 0.001 & 2.52 \\
	Fe I & 15490.381 & -0.004 & 0.001 & -2.722 \\
	Fe I & 15692.643 & -0.005 & 0.002 & -2.759 \\
	Fe I & 16009.512 & -0.006 & 0.003 & -2.143 \\
	Si I & 15960.044 & -0.007 & 0.003 & -2.112 \\
	Si I & 16680.77 & -0.006 & 0.003 & -2.156 \\
	K I & 15163.033 & 0.006 & 0.003 & 2.043 \\
	V I & 15923.924 & 0.002 & 0.001 & 2.017 \\
	Cr I & 15680.073 & -0.005 & 0.002 & -2.811 \\
	Mn I & 15159.054 & -0.008 & 0.002 & -3.298 \\
	Mn I & 15262.023 & -0.005 & 0.002 & -2.996 \\
	\enddata
\end{deluxetable} \label{table:feh_derivative}
\FloatBarrier

The spectra contain roughly 8000 pixels, so we might
expect the $\chi^2$ values to be close to 8000 in magnitude, but
they are much higher. This discrepancy follows from the fact
that, while the spectral model is good at the level of a few per
cent, the signal-to-noise ratio of a typical spectrum is more
than 100. That is, the $\chi^2$ values do show that the model is not
good in the frequentist sense; it is only good at the level of a
few per cent.

\begin{figure}
	\begin{center}
	\includegraphics[width=\linewidth]{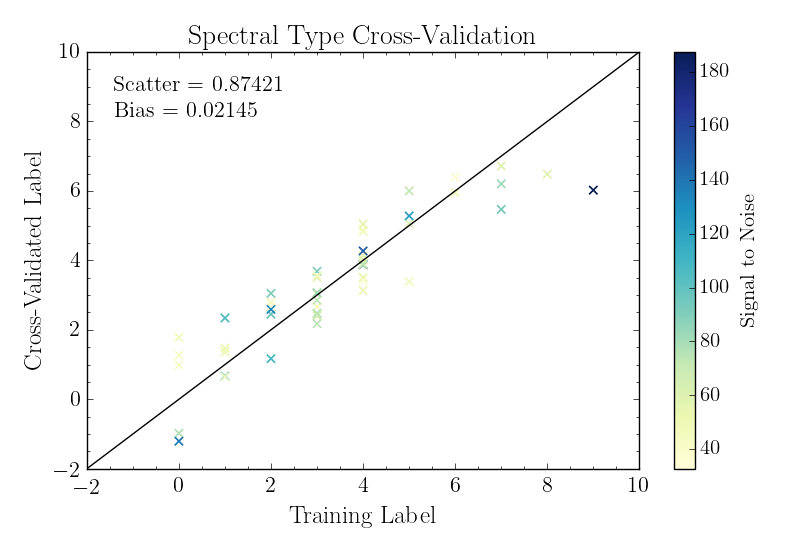}
	\end{center}
	\caption{Leave-one-out cross-validation test for the West-trained spectral type model. Predictive accuracy, as computed from the scatter in cross-validation is 0.9 subtypes.}
	\label{fig:west_validation}
\end{figure}

\subsection{Spectral Type Model \label{subsec:west_results}}

We trained \thecannon\ on 51 M dwarfs in the range M0$-$M9 with a one-dimensional spectral type label, and obtained a precision of $\pm$0.9 spectral types, similar to the uncertainty of
the original training label of $\pm1$ spectral type. We note,
however, that the training sample is distributed heavily toward
sources of earlier type, with a median spectral type of 3 and
only one M8 and one M9 source. As seen in Figure \ref{fig:west_validation}, the model performs poorly at reproducing spectral types $>$M8, which confirms that \thecannon\ does not extrapolate well to labels outside of the training sample space. Because of this skew for late-type sources, we report our spectral type model to be precise to $\pm$0.7 spectral types for the range M0-M6. Repeating the analysis of Section \ref{subsec:mann_results}, Figure \ref{fig:west_validation} shows LOOCV test for the labels reported in Table \ref{table:west_results}, and Figure \ref{fig:west_derivative} shows the derivative of model flux with varying spectral type.

\begin{deluxetable*}{ccc|ccc|c}
\tablecaption{Cannon results for Spectral Type model. Reported uncertainties for the W11 training sample labels are $\pm$1 spectral types, and reported uncertainties for \thecannon\ model based on the cross-validation scatter are $\pm$0.7 spectral types for types $\leq$M6. \label{table:west_results}}
\tabletypesize{\scriptsize}
\tablehead{
\multicolumn{3}{c}{\underline{Designation}} & \multicolumn{3}{c}{\underline{Spectral Type}}
	& \multicolumn{1}{c}{\underline{Model Fit}} \\
	  \colhead{2MASS ID} & \colhead{RA}   & \colhead{DEC}
    & \colhead{Training} & \colhead{Test} & \colhead{LOOCV} 
    & \colhead{Test $\chi^2$}
} 
	\startdata
	2M03114974+0115158 & 47.957262  & 1.254404  & 1          & 0.9       & 1.4        & 78236   \\
	2M03122509+0021585 & 48.104563  & 0.366251  & 7          & 7.3       & 5.5        & 56504   \\
	2M03423963+0012102 & 55.66513   & 0.202859  & 4          & 3.9       & 3.9        & 29598   \\
	2M04262170+1800009 & 66.590421  & 18.000265 & 5          & 5.0       & 5.1        & 14084   \\
	2M09152918+4407461 & 138.871602 & 44.129498 & 3          & 3.6       & 3.7        & 17921   \\
	2M09183649+2207022 & 139.652051 & 22.117298 & 3          & 2.4       & 2.4        & 10591   \\
	2M09332262+2749021 & 143.344279 & 27.817253 & 3          & 2.8       & 2.7        & 17388   \\
	2M09373349+5534057 & 144.389577 & 55.568275 & 6          & 6.5       & 6.4        & 15335   \\
	2M10313413+3441535 & 157.892222 & 34.698212 & 1          & 2.0       & 2.3        & 21134   \\
	2M11194647+0820356 & 169.943658 & 8.343246  & 8          & 7.5       & 6.5        & 37803   \\
	2M11203609+0704135 & 170.1504   & 7.070432  & 6          & 6.1       & 5.9        & 18677   \\
	2M11570299+2028436 & 179.262465 & 20.4788   & 2          & 2.9       & 3.0        & 16205   \\
	2M12203634+2505351 & 185.15143  & 25.093107 & 4          & 3.4       & 3.1        & 85745   \\
	2M12212701-0030560 & 185.362567 & -0.515566 & 0          & 1.4       & 1.8        & 22646   \\
	2M12423245-0646077 & 190.635249 & -6.768827 & 2          & 2.4       & 2.5        & 18410   \\
	2M12464541-0312524 & 191.689212 & -3.214578 & 4          & 3.6       & 3.5        & 11561   \\
	2M12471099+1109566 & 191.795795 & 11.165737 & 3          & 3.1       & 3.0        & 12738   \\
	2M12492657-0312032 & 192.360749 & -3.200903 & 3          & 3.3       & 3.5        & 37756   \\
	2M12503440+4309482 & 192.643353 & 43.163414 & 4          & 4.0       & 4.0        & 19726   \\
	2M12523816+1240586 & 193.159002 & 12.682945 & 4          & 4.1       & 4.1        & 15526   \\
	2M12552141+4150425 & 193.839213 & 41.845161 & 3          & 3.1       & 3.1        & 29274   \\
	2M12564117+4233175 & 194.171558 & 42.554871 & 3          & 3.1       & 3.1        & 35083   \\
	2M13032161+4220407 & 195.840051 & 42.344654 & 2          & 2.7       & 2.8        & 14657   \\
	2M13415860+1852278 & 205.494169 & 18.874393 & 4          & 3.9       & 3.9        & 9475    \\
	2M13442970+5625445 & 206.123779 & 56.429039 & 5          & 5.9       & 6.0        & 16017   \\
	\nodata & \nodata & \nodata & \nodata & \nodata & \nodata & \nodata
	\enddata
\end{deluxetable*}


\begin{figure*}
	\includegraphics[width=\linewidth]{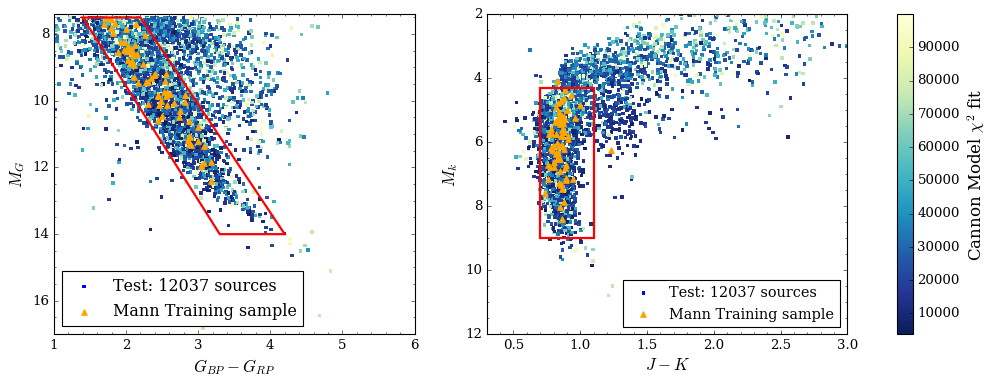}
	\caption{\gaia\ and 2MASS color-magnitude cuts for the 12,037 sources with $\chi^2<100,000$. Overplotted with orange triangles are the 67 out of 87 training sample sources that have parallaxes measured by \gaia. The coordinates for the selected quadrangles are
	$\left\{ (1.4, 7.5), (2.2, 7.5), (4.2, 14), (3.3, 14) \right\}$ corresponding to ($BP-RP$, $M_G$) for the \gaia\ color-magnitudes shown in the left panel, and 
	$\left\{ (.7, 4.3), (1.1, 4.3), (.7, 9), (1.1, 9) \right\}$ corresponding to ($J-K$, $M_K$) for the 2MASS color-magnitudes shown in the right panel. }
	\label{fig:cmd_selection}
\end{figure*}

\subsection{Test Sample \label{subsec:test_selection}} 

\begin{figure}
\begin{center}
	\includegraphics[width=\linewidth]{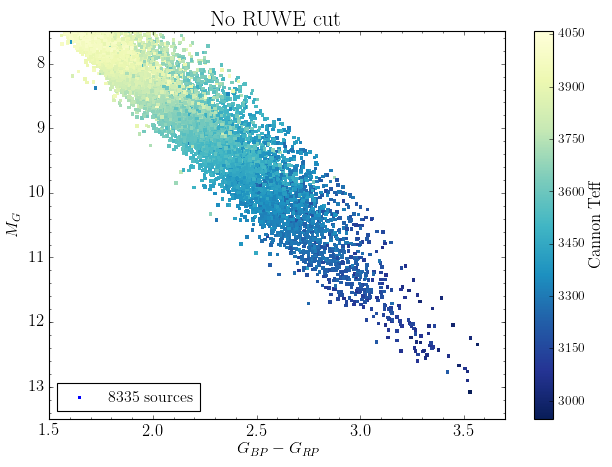} 
	\includegraphics[width=\linewidth]{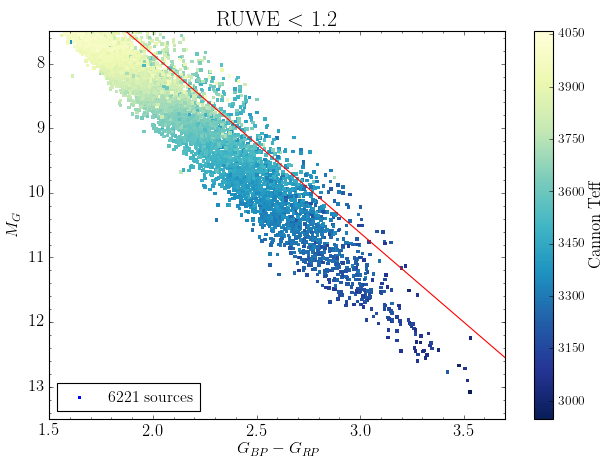}
	\caption{The top panel shows the test sample of 8335 sources after applying Selections $1-3$ described in Section \ref{subsec:test_selection}. The bottom panel shows the same test sample reduced to 6221 after applying an astrometric quality cut of ${\rm RUWE}<1.2$ (Selection 4). To further remove sources that were likely binary contamination (Selection 5), we cut out sources above the red line that sparsely lay above the majority of the main-sequence with temperatures and metallicities that deviate from the expected gradient, reducing the final test sample to a size of 5875 sources. The red line shown is constrained by the ($BP-RP$, $M_G$) coordinates $\left\{ (3.5,12), (1.87,7.5) \right\}$.}
	\label{fig:ruwe_cut}
\end{center}
\end{figure}

Out of the total APOGEE DR14 catalog of 258,475 sources, we selected 254,478 sources that were in the cross-match of \drtwo\ \citep{Brown:2018} and applied \gaia\ color-magnitude cuts of $1<BP-RP<6$ and $7.5<M_G<20$ for sources with only positive parallaxes ($\varpi>0$), yielding a sample of 14,828 sources. From there we applied additional selection criteria, described below, to identify a sample of single, main-sequence M stars, with minimal contamination from reddened K dwarfs, pre-main-sequence stars, and binaries:

\begin{enumerate}
\item \textit{Quality of fit cut:} We apply a \cannon\ model $\chi^2$ cut of less than 100,000, chosen to remove badly fit sources (such as fast rotators), but include $\chi^2$ values close to the distribution of training sample (Figure \ref{fig:chi_dist}).

\item \textit{Color-magnitude cuts:} Using \gaia\ and photometry from
the Two Micron All Sky Survey (2MASS) we apply the additional color-magnitude selections shown in Figure \ref{fig:cmd_selection} to remove sources above the main sequence (which are likely pre-main-sequence, reddened K dwarfs and/or multiples), and subdwarfs below the main sequence.

\item \textit{Model extrapolation cuts:} Because \thecannon\ does not perform well under extrapolated regions of parameter space, we select only sources inside the range of our training sample with ASPCAP parameters $2800<T_{\rm eff}<4100$ and $-0.5<\mh<0.5$, and with \cannon\ parameters of $2850<T_{\rm eff}<4150$, $-0.5<\feh<0.5$, and $0<SPT<9$.

\item \textit{Astrometric cut:} Using the \gaia\ renormalised unit weight error (RUWE)--a metric of evaluating the fit of the astrometric solution described in the additional release notes \citep{Lindegren:2018}--we apply a cut of ${\rm RUWE}<1.2$ to remove sources with high astrometric error or noise, such as binaries (see Figure \ref{fig:ruwe_cut}).

\item \textit{Binary cut:} To remove further contamination from binary sources, we applied an additional color-magnitude cut on sources above the main sequence, which we visibly selected for in Figure \ref{fig:ruwe_cut}.

\end{enumerate}

\begin{figure*}
\begin{center}
	\includegraphics[width=.47\linewidth]{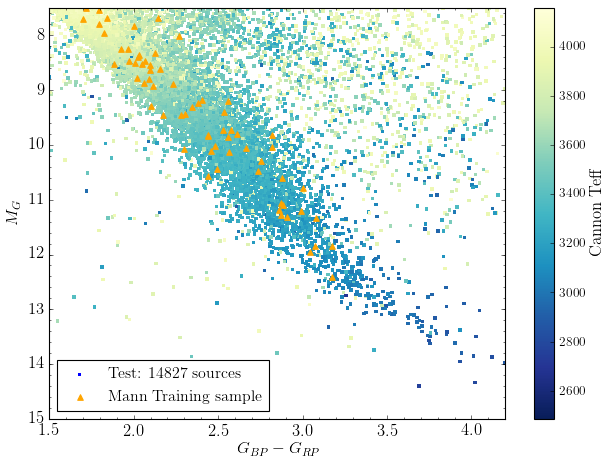}
	\includegraphics[width=.47\linewidth]{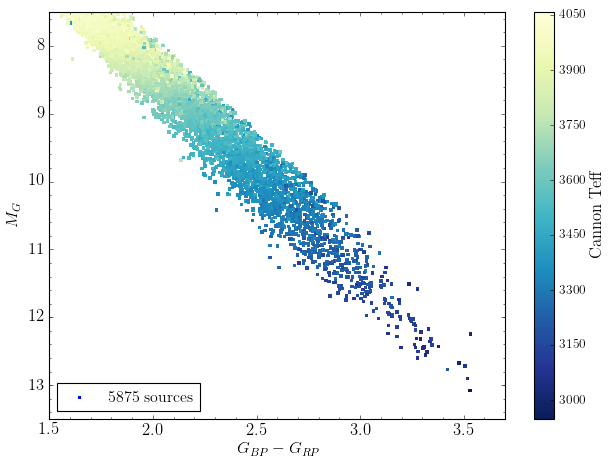} \\
	\includegraphics[width=.47\linewidth]{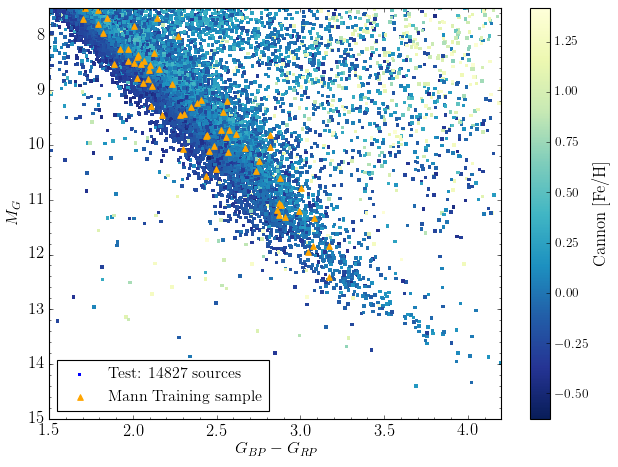}
	\includegraphics[width=.47\linewidth]{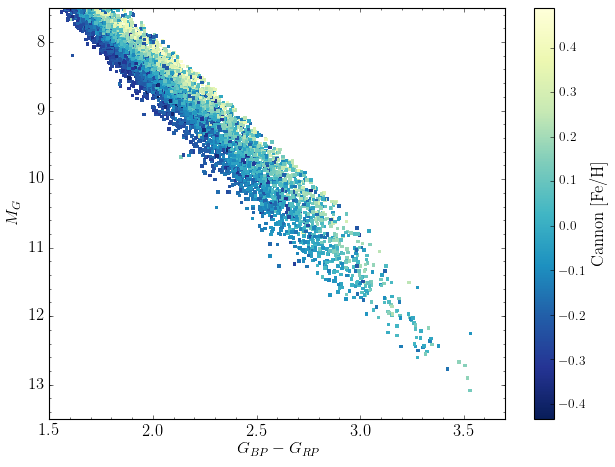} \\
	\includegraphics[width=.47\linewidth]{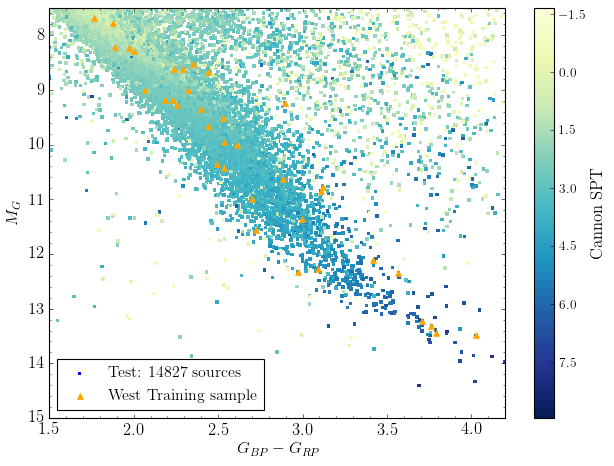}
	\includegraphics[width=.47\linewidth]{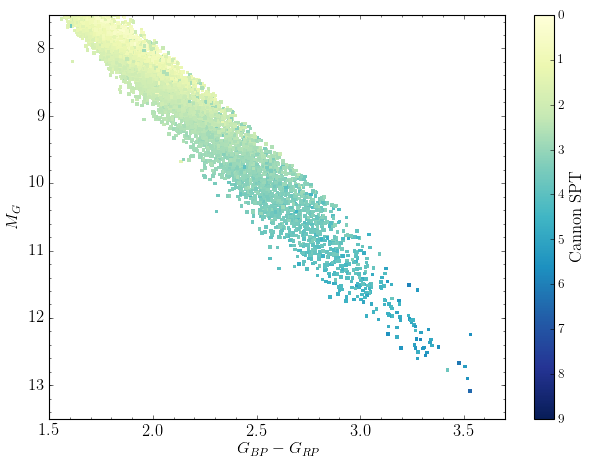}
	\caption{Full sample of 14,828 M dwarfs colored by \cannon\ labels before selection (left) and final sample selection of 5,875 M dwarfs after applying selection criteria described in Section \ref{subsec:test_selection} (right), to reduce contamination from sources with that are not similar to the training sample (not single, main-sequence M stars, such as pre-main sequence, spectroscopic binaries, and K dwarfs). Overplotted with orange triangles are the M15 and W11 training samples, for their respective \cannon\ test labels. Temperature gradient increases with decreasing color, spectral subtype increases with increasing color, and metallicity gradient increases perpendicularly up from the main sequence branch as expected. Deviations from these gradients seen at the upper boundary of the main sequence is likely remaining contamination from the binary sequence.}
	\label{fig:safe_selection}
\end{center}
\end{figure*}

The top, middle, and bottom panels of Figure \ref{fig:safe_selection} show before and after selection of the sources in \gaia\ color-magnitude space, colored by temperatures, metallicities, and spectral types determined by \thecannon, with their respective training samples overplotted in orange. Each plot shows the expected gradient: temperature increases with decreasing color, spectral subtype increases with increasing color, and the metallicity gradient is largely perpendicular to the main-sequence branch. 
We also note that applying our model requires very little computational demand: the time to train and test a model on all 14,828 sources was two minutes on a 2.7\,GHz Intel core i7 laptop. 

Table \ref{table:test_results} outlines the parameters included in the test sample catalog, which can be downloaded from the online journal. Included are two versions of the catalog: the first containing all 14,828 sources before selection, and the second containing the 5875 sources kept after making selections $1-5$ described in this section.

\subsection{Temperature Validation}

\begin{figure*}
\begin{center}
	\includegraphics[width=.9\linewidth]{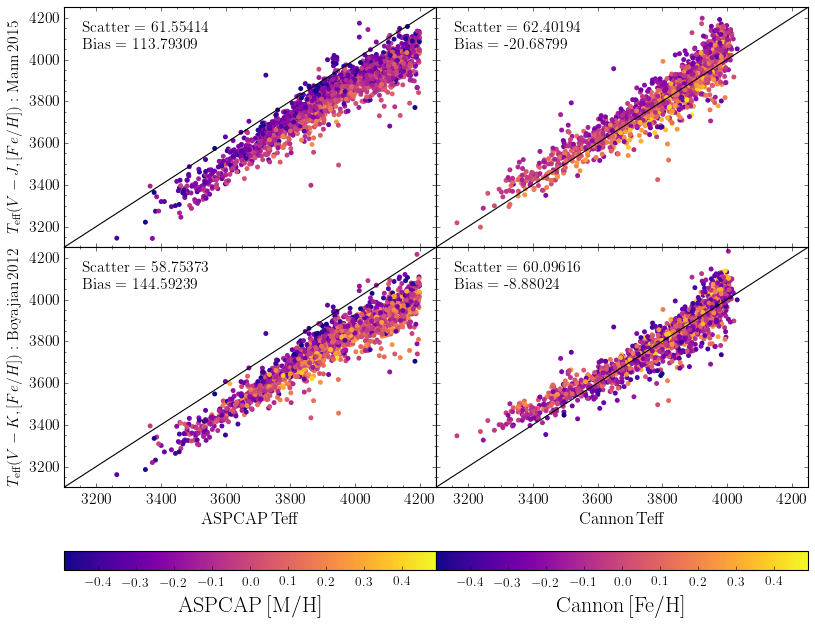}
	\caption{\cannon\ (right column) and ASPCAP DR14 (left column) temperatures as compared to temperatures estimated color-temperature relations from three different literature sources for a subsample of 1702 sources with $V,J,K$ colors. Shown are \cannon\ and ASPCAP temperatures (x-axis) compared to temperatures estimated from the $V-J$ and $\feh$ dependent relation of \citealt{Mann:2015} (top), and the relation between $V-K$ and $\feh$ of \citealt{Boyajian:2012} (bottom).}
	\label{fig:teff_comparisons}
\end{center}
\end{figure*}

As a validation test of our derived temperatures, we perform a comparison between several color-temperature relations from literature, which use combinations of 2MASS and visual band photometries to predict temperatures (similarly to the evaluation of ASPCAP temperatures by \citealt{Schmidt:2016}). To obtain visual-band magnitudes for a set of sources, we cross-matched the 5875 sources in our ``safe'' test sample to the AAVSO Photometric All-Sky Survey DR9 (APASS; \citealt{Henden:2016}), to obtain a subsample of 1702 sources with both $BV$ photometries measured by APASS and 2MASS $JHK$ photometries from APOGEE. Figure \ref{fig:teff_comparisons} shows \cannon\ vs. photometric temperatures on the right, and ASPCAP vs. photometric temperatures on the left for each of the 1702 sources, colored by their respective spectroscopic metallicities.

Compared to the \citet{Mann:2015} and \citet{Boyajian:2012} color-metallicity-derived temperatures, both ASPCAP and \cannon\ temperatures show similar scatters of $\sim60\,$K, but are offset by a constant.
We find \cannon\ to be in better agreement with \citet{Mann:2015} and \citet{Boyajian:2012}, with ASPCAP overestimating $\teff$ on average by $\sim110-140\,$K, and \thecannon\ underestimating $\teff$ on average by $\sim10-20\,$K, with the largest deviation in the latter at the lowest and highest $\teff$.

\subsection{Metallicity Validation}

\begin{figure}
\begin{center}
	\includegraphics[width=\linewidth]{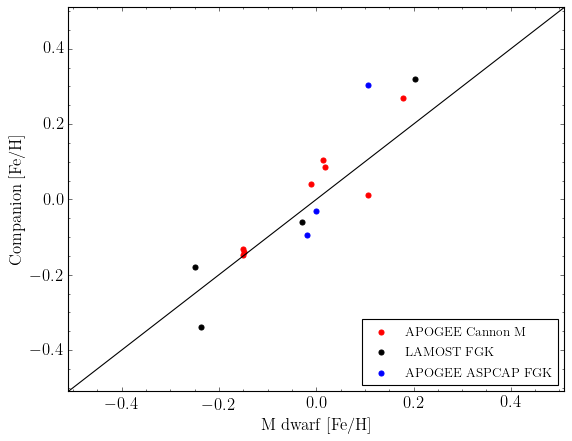}
	\caption{Metallicities of \cannon\ M dwarfs in binaries (x-axis) compared to the metallicities of their companions (y-axis). Label colors show the source and spectral type of the companion metallicities. Metallicity values for each binary pair are compiled in Table \ref{table:binaries}. The overall scatter between the 15 metallicity pairs is 0.08\,dex.} 
	\label{fig:wide_binary_met}
\end{center}
\end{figure}

As a check of the reliability of our test sample metallicity,
we cross-matched our M-dwarf final sample with the catalog of
$>$50,000 high-confidence, widely separated binaries identified
by \drtwo presented in \citet{Elbadry:2018b}.
In total we found 216 of the APOGEE M dwarfs to have binary pairs (46 FGK+M, 155 M+M, and 15 WD+M). Out of the 155 M+M pairs, eight pairs contained both pairs in APOGEE. Cross-matching the list of FGK+M dwarf companions with several catalogs/surveys with measured stellar metallicities, we found an additional seven sources with FGK metallicities from LAMOST \citep{Zhao:2012} and APOGEE (ASPCAP). The metallicity measurements for the 15 M-dwarf binaries and their companions are given in Table \ref{table:binaries} and shown in Figure \ref{fig:wide_binary_met}, and the overall scatter is $0.08\,$dex--an improvement over the scatter of ASPCAP metallicities, which is $0.15\,$dex for these 15 sources. The internal consistency of the two models (the scatter of the eight M+M pairs both in APOGEE) is 0.06\,dex for \cannon\ and 0.12\,dex for ASPCAP.

\begin{figure*}
\begin{center}
	\includegraphics[height=5cm]{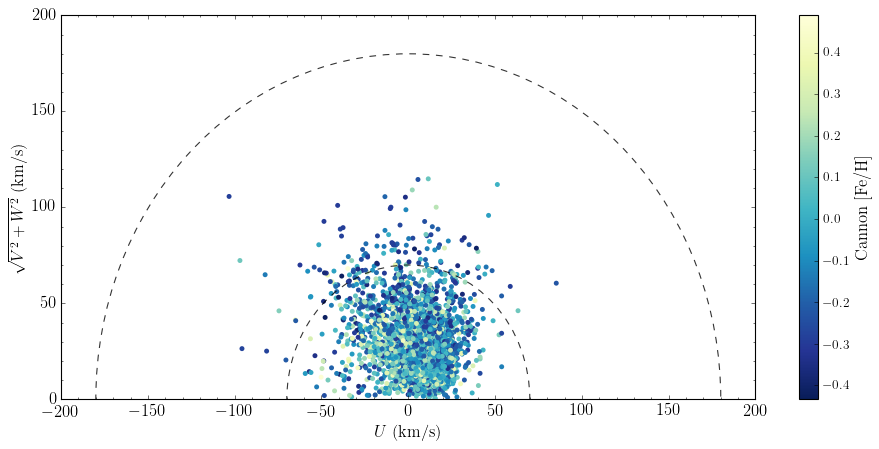}
	\includegraphics[height=5cm]{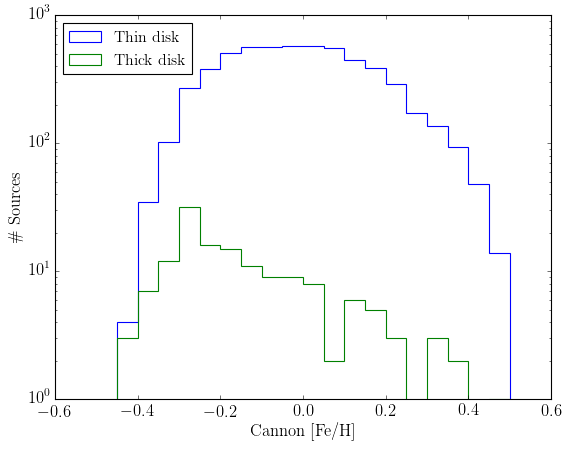}
	\caption{Toomre diagram showing the distribution of the \cannon\ metallicities (left). Dashed lines at $v_{tot}=70$\,km/s and 180\,km/s show roughly the separation of thin/thick/halo stars. The histogram (right) shows the metallicity distribution of thin- and thick-disk stars, with blue corresponding to $v_{tot}<70$\,km/s and green corresponding to $70<v_{tot}<180$\,km/s respectively. The average/standard deviation metallicity of the thin-disk distribution is $\feh=0.00\pm0.17\,$dex, and that of the thick-disk distribution is $\feh=-0.14\pm0.19\,$dex.}
	\label{fig:toomre}
\end{center}
\end{figure*}

As expected, the Toomre diagram in Figure \ref{fig:toomre} shows that higher-metallicity sources in the sample are concentrated in low-velocity space corresponding roughly to the thin disk population; while the thick-disk population contains a slightly higher concentration of lower-metallicity sources. Separating the two populations into separate histograms (also shown in Figure \ref{fig:toomre}), we find that thick-disk sources are marginally more metal poor than thin-disk sources, with the mean $\pm$ standard deviation of $\feh=0.00\pm0.17\,$dex for the thin-disk distribution, and $\feh=-0.14\pm0.19\,$dex for the thick-disk distribution.
Metallicities of the two populations from ASPCAP show a similar distribution, with the mean $\pm$ standard deviation being $\mh=-0.16\pm0.16\,$dex for thin-disk sources and $\mh=-0.23\pm0.17\,$dex for thick-disk sources.

\begin{figure*}
\begin{center}
	\includegraphics[width=.3\linewidth]{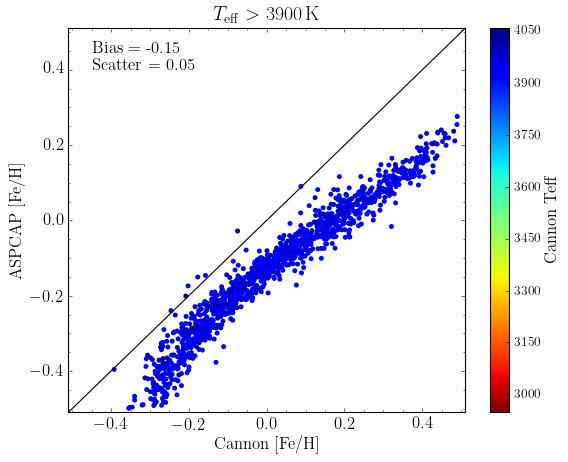}
	\includegraphics[width=.3\linewidth]{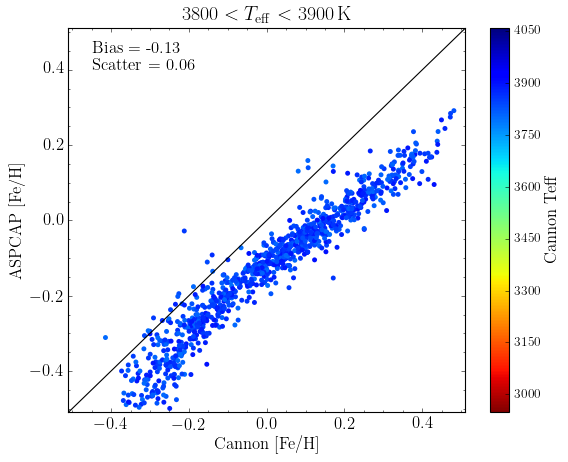}
	\includegraphics[width=.3\linewidth]{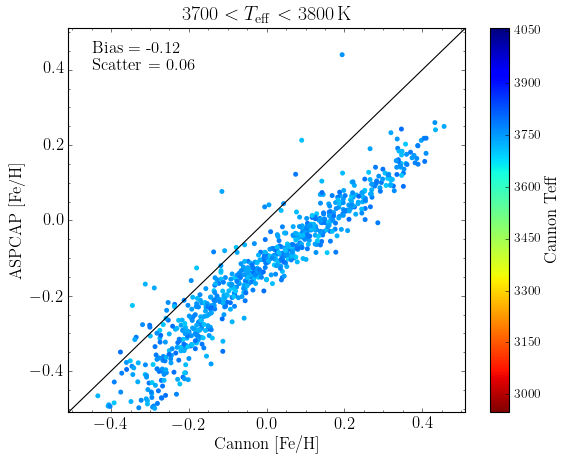} 
	\includegraphics[width=.3\linewidth]{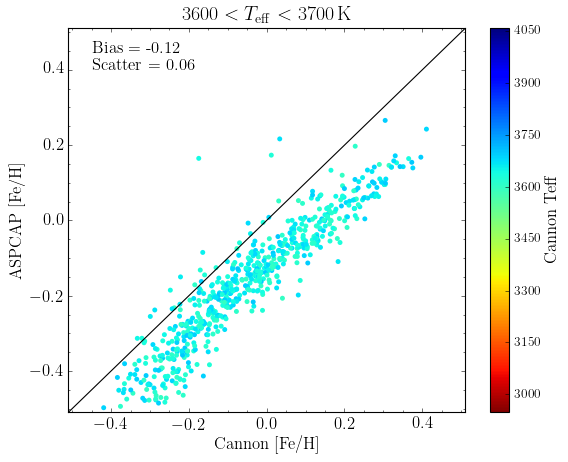}
	\includegraphics[width=.3\linewidth]{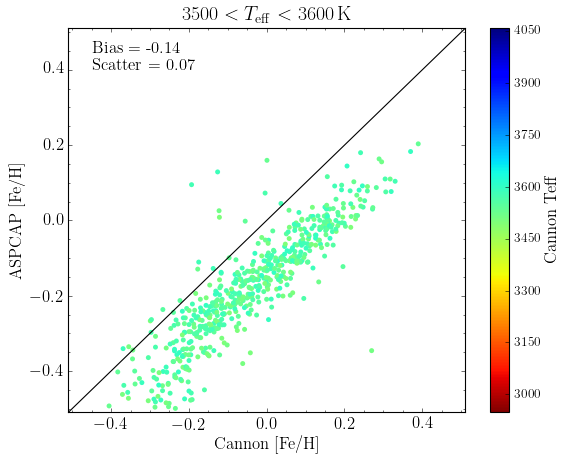}
	\includegraphics[width=.3\linewidth]{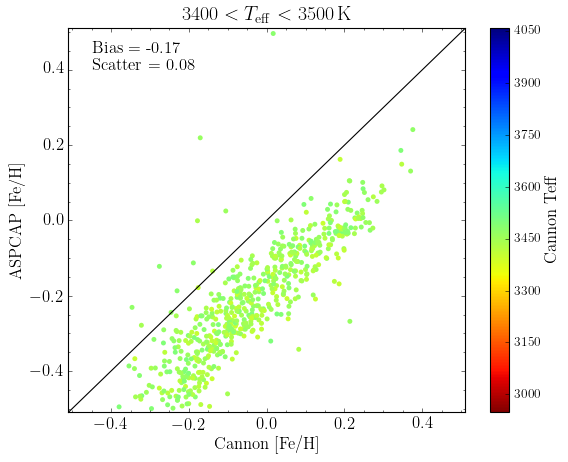} 
	\includegraphics[width=.3\linewidth]{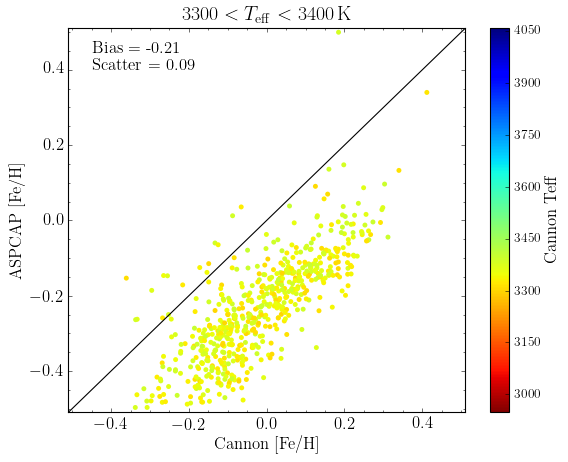}
	\includegraphics[width=.3\linewidth]{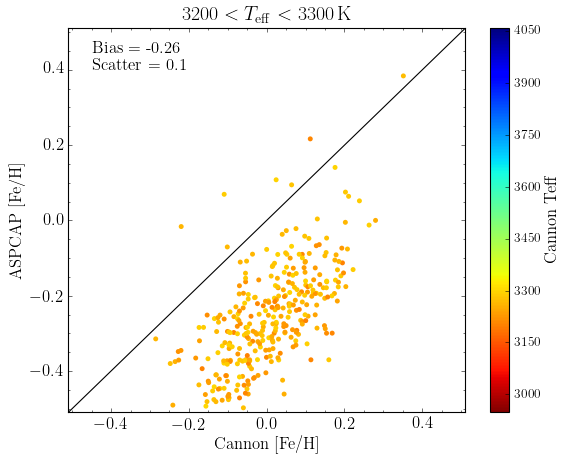}
	\includegraphics[width=.3\linewidth]{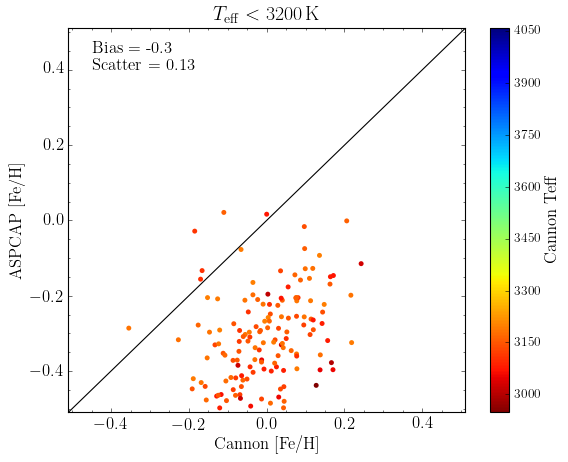} 
	\caption{A comparison of ASPCAP DR14 and Cannon metallicities for 8335 test sample sources separated into temperature bins of 100K. We see that the scatter and metal-poor bias of ASPCAP DR14 metallicities clearly increases with decreasing temperature.} 
	\label{fig:met_comp}
\end{center}
\end{figure*}

Figure \ref{fig:met_comp} shows that ASPCAP metallicities are systematically lower than \cannon\ metallicities. We further find that the bias is temperature-dependent: at the highest temperatures ($\teff>3600\,$K) ASPCAP and \cannon\ metallicities are consistent to a scatter of $0.05-0.06\,$dex and offset by an average of $-0.12-0.15\,$dex, while at the lowest temperatures ($\teff<3200\,$K) ASPCAP and \cannon\ are consistent to a scatter of $\sim-0.13\,$dex and offset by an average of $-0.3\,$dex.


\section{Discussion} \label{sec:discussion}

We trained a data-driven model (\thecannon; \citealt{Ness:2015}) to deliver
high-quality atmospheric parameters ($\teff$ and $\feh$) for M-type dwarf stars
from high-resolution infrared spectra from APOGEE.
This work was motivated by the problem that M dwarfs stars are difficult to
model physically; the data are better than the models in important senses.
Indeed we find that our data-driven model is both accurate in the
data domain (as a spectral synthesis model) and precise in the latent domain
(as a tool for deriving physical parameters).
This accuracy and precision is consistent with previous work with
\thecannon\ (\citealt{Ness:2015, Casey:2016, Ho:2017a, Ness:2018}), but
here extends to a new regime in spectral type ($\teff$).
The primary result of this work is that
we have compiled a catalog of 5875 M dwarfs with \cannon\ temperatures,
metallicities, spectral types, and six-dimensional kinematics.
These data are provided in Table \ref{table:test_results}.

While \thecannon\ achieves excellent precision at predicting labels
and reproducing spectral features, the
accuracy of labels it produces is limited by
the accuracy, relative precision, size, dynamic range, and
representation of the training sample.
That is, being a supervised method, \thecannon\ is never any better in a mean (bias) sense than the input training data, although it can be better in a precision or variance sense.
The catalog we have produced is a label transfer from parameters provided
in our input data (M15) and it implicitly adopts all the biases and
issues from those input data.
It is also limited to the stellar-parameter domain of that input catalog.
That said, this work provides an external validation of the M15 stellar parameters.

The model we have developed does have limitations, however.
For example, it delivers chi-squared goodness-of-fit measures that
are large; the model is not technically an accurate description of the spectra,
especially when the spectra are observed at signal-to-noise levels above 100.
The model does not include some known physical and instrumental effects.
such as line broadening from rotation or
convection (for example, \citealt{Behmard2019}), 
or binarity and the superposition of multiple
stellar spectra (as in, say, \citealt{ElBadry:2018}).
The model also does not include any adjustments for instrumental variations, such as 
the small but significant variations of APOGEE resolution with
spectrograph fiber number (as included in \citealt{Ness2018}).

The APOGEE instrument was designed to be sensitive to more than
a dozen individual elemental abundances in stellar spectra.
So the M-dwarf spectra analyzed here contain individual elemental abundance
information that we have ignored.
Exploitation of that information requires a better training set of M dwarfs than
we have at present, but is an important goal for the future with these data.

While a detailed analysis of atmospheric model limitations is
beyond the scope of this paper, our results provide an avenue to
compare the metallicity scale for FGK stars to the less well-
understood metallicity scale for M dwarfs.
These results find that atmospheric metallicities are systematically metal-poor biased
compared to \cannon-based metallicities trained on sources with metallicities calibrated
to those of FGK companions. 
At the high-temperature end ($\teff > 3600$\,K), the ASPCAP metallicity bias is $-0.12-0.15\,$dex with a scatter of $0.05-0.06\,$dex relative to \cannon\ metallicities, and it increases to a bias of $-0.3\,$dex and scatter of $0.13\,$dex at the lower temperature end ($\teff < 3200$\,K) (Figure \ref{fig:met_comp}).
We suspect that this metal-poor bias, while not explored to a great extent in this work, is due 
to the line lists of the models--an effect in which the optimizer of the pipeline may be lowering the continuum level and metallicity of the fit to compensate for the missing lines or opacities.
We also note that this analysis was completed using data from Data Release 14 of APOGEE, 
which did include molecular lines from FeH in the pipeline at the time, which become numerous and strong for $\teff \lesssim 3600$\,K \citep{Souto:2017}.
Further analysis would
need to be done to quantify the metallicity improvement for M
dwarfs in future data releases of APOGEE, and determine
whether the metallicity bias is found in other model grids
(besides the ATLAS/MARCS models used by the ASPCAP DR14
pipeline), and whether the effect is present at other
wavelengths.

\begin{figure*}
	\begin{center}
	\includegraphics[width=16cm]{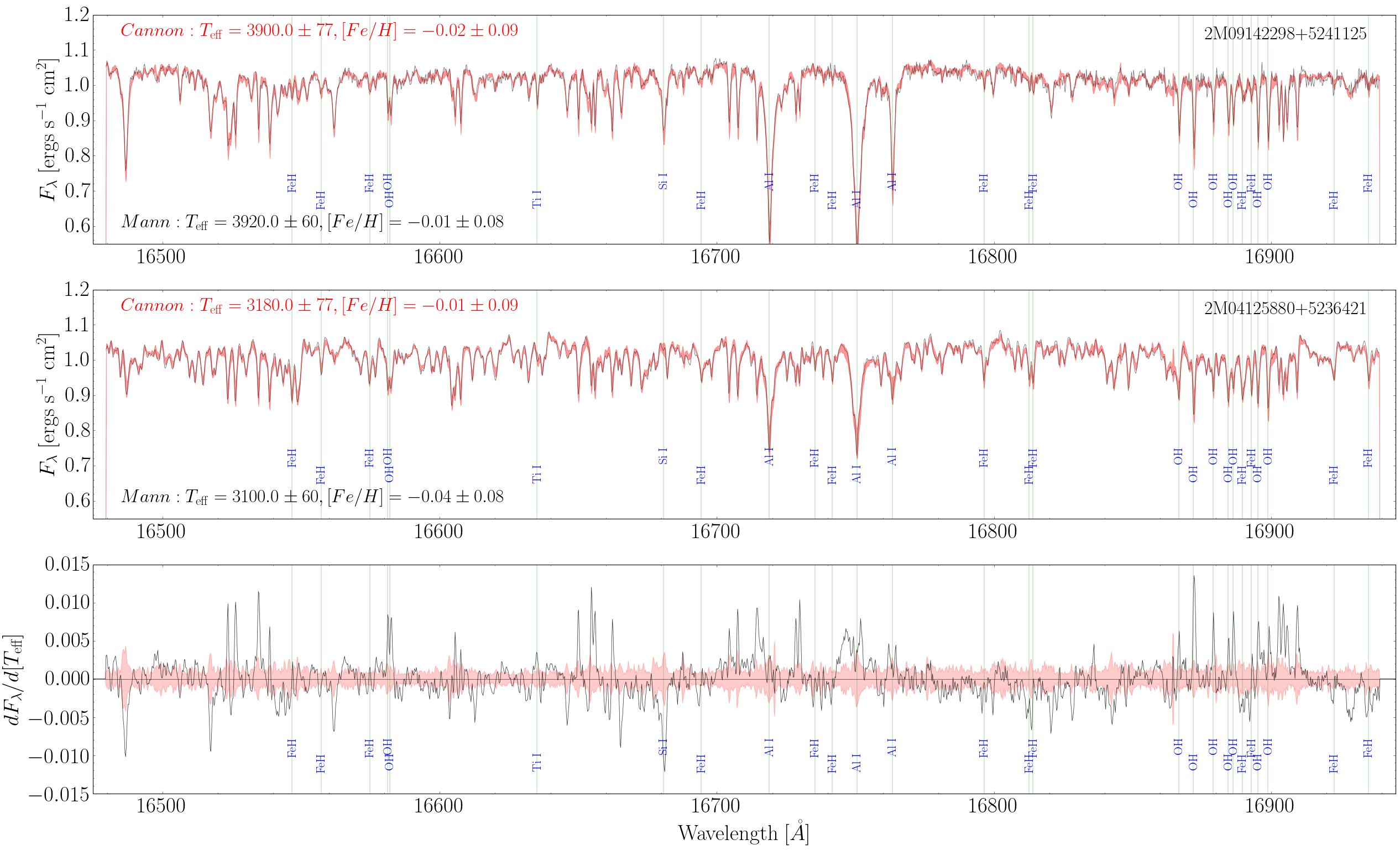}
	\end{center}
	\caption{\textit{Top two panels of each plot:} APOGEE spectra (black), overlaid by the Mann-trained \cannon\ model for two sources of varying temperatures, and similar metallicities. \textit{Third panel of each plot:} Derivative of \thecannon\ model with respect to temperature, taken at the median training temperature, T$_{\rm eff}=3463$K; an error estimate computed using a jackknife statistic at each pixel is marked in red, making it possible to distinguish which features vary significantly with change in spectral type, and which are likely due to noise.}
	\label{fig:demo_teff}
\end{figure*}

\begin{figure*}
	\begin{center}
	\includegraphics[width=16cm]{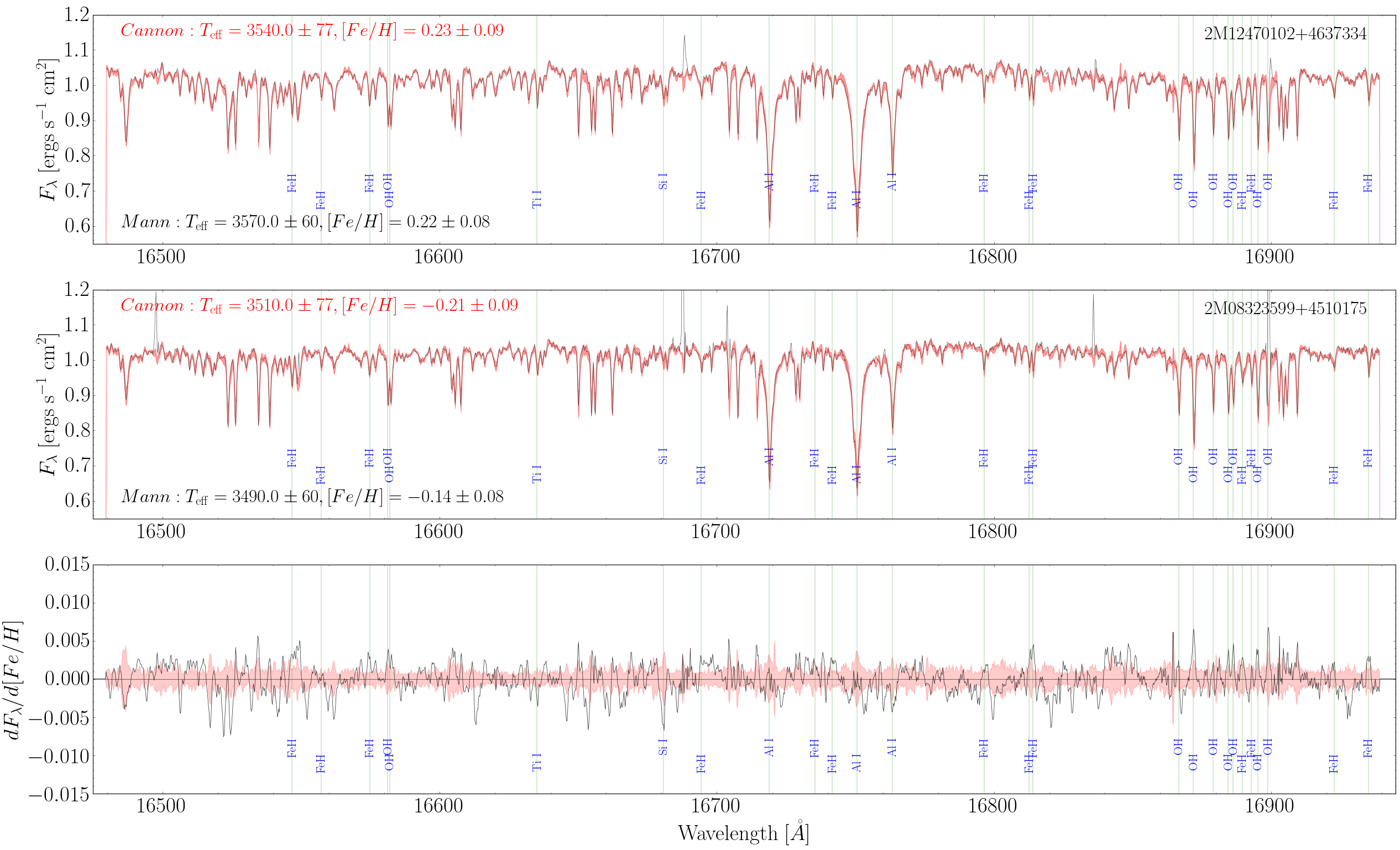}
	\end{center}
	\caption{\textit{Top two panels of each plot:} APOGEE spectra (black), overlaid by the Mann-trained \cannon\ model for two sources of varying metallicities and similar temperatures. \textit{Third panel:} Derivative of \thecannon\ model with respect to metallicity, taken at the median training metallicity, $\feh=-0.03$\,dex; the jackknife computed error at each pixel is shown in red.} \label{fig:demo_feh}
\end{figure*}

\begin{figure*}
	\begin{center}
	\includegraphics[width=16cm]{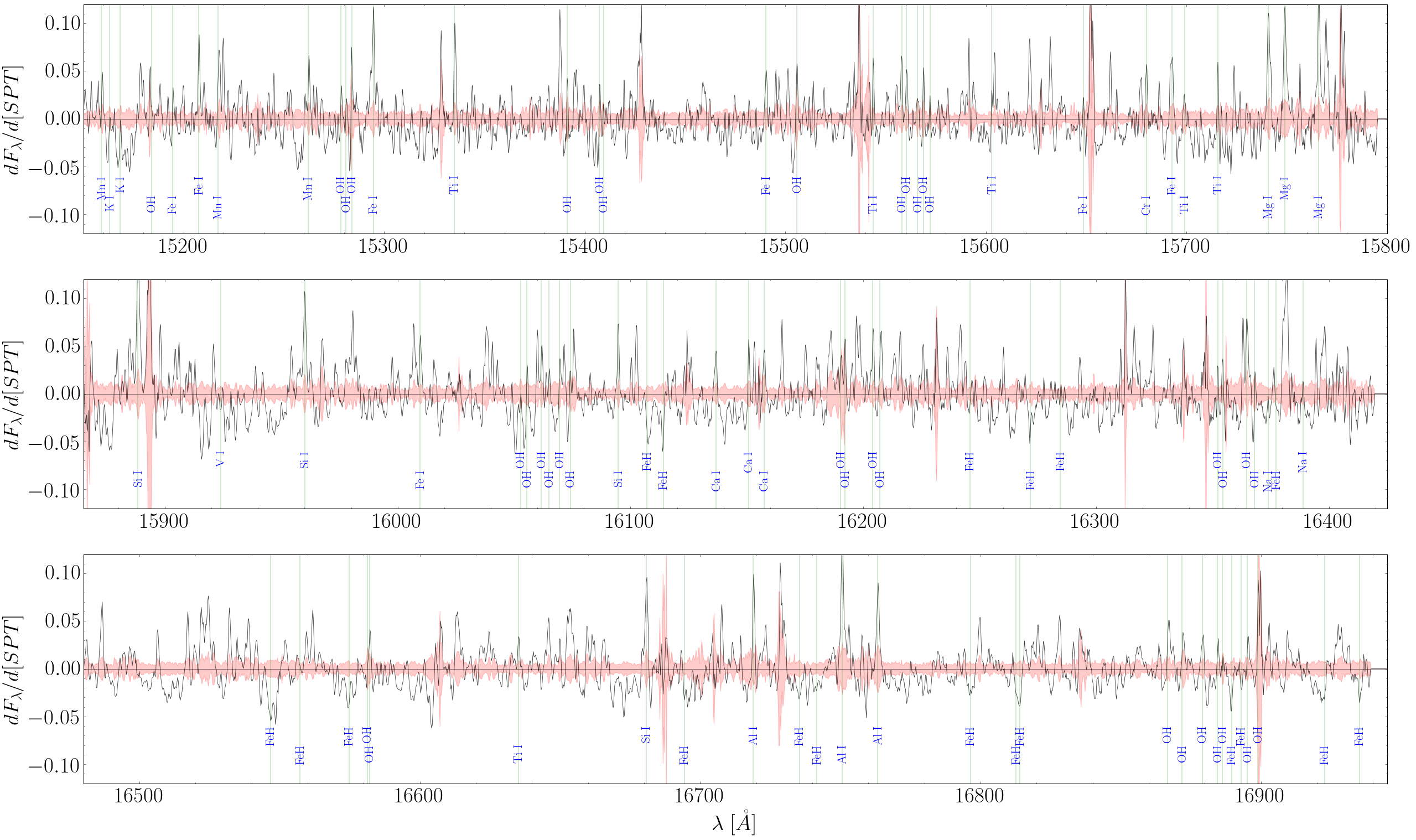}
	\end{center}
	\caption{Derivative plots for the spectral type model taken at the median training spectral type, SPT$=3$; the jackknife computed error at each pixel is shown in red.} 
	\label{fig:west_derivative}
\end{figure*}

Given that physics-based spectral models of M dwarfs have issues, one of the possible
future values of the data-driven model shown here is that it is highly interpretable:
it contains within it first and second derivatives of the spectral expectation with
respect to the atmospheric parameters.
We show some of these derivatives in Figures \ref{fig:demo_teff}, \ref{fig:demo_feh}, and \ref{fig:west_derivative} and deliver relevant data in
Table~\ref{table:teff_derivative} and Table~\ref{table:feh_derivative}.
These tables summarize spectral features in the APOGEE bandpass
that are found to be strong temperature and metallicity indicators.
In the long run, this is the primary value of data-driven models for astronomy:
to provide physical insights that drive physical understandings.
It is our hope that \thecannon, and models like it, will lead to
new and improved physical models which will, in turn, put \thecannon\ out of business. 

\acknowledgements
We would like to acknowledge Hans Walter Rix (MPIA), Derek Homeier (Universit{\"a}t Heidelberg), Wolfgang Bradner (MPIA), Anna-Christina Eilers (MPIA), Melissa Ness (Columbia), Kevin Covey (WWU), Diogo Souto (Observatário Nacional/MCTI), Keivan Stassun (Vanderbilt), Katia Cunha (NOAO), Anthony Brown (Leiden), Aida Behmard (Caltech) and Christopher Theissen (UCSD) for constructive discussions in the process of this project, as well as Bertrand Goldman (MPIA) and those who have supported the internship program at the Max Planck Institute f{\"u}r Astronomie for providing JB with funding and hospitality.
This project was partially supported by
the US National Aeronautics and Space Administration (NASA grant NNX12AI50G),
the US National Science Foundation (NSF grant AST-1517237),
and the Moore--Sloan Data Science Environment at NYU.
A.J.B. acknowledges funding support from the National Science Foundation under award No. AST-1517177.
This work is supported by the SDSS Faculty and Student Team (FAST) initiative.

Funding for the Sloan Digital Sky Survey IV has been provided by the Alfred P. Sloan Foundation, the U.S. Department of Energy Office of Science, and the Participating Institutions. SDSS-IV acknowledges support and resources from the Center for High-Performance Computing at
the University of Utah. The SDSS web site is www.sdss.org.

SDSS-IV is managed by the Astrophysical Research Consortium for the 
Participating Institutions of the SDSS Collaboration including the 
Brazilian Participation Group, the Carnegie Institution for Science, 
Carnegie Mellon University, the Chilean Participation Group, the French Participation Group, Harvard-Smithsonian Center for Astrophysics, 
Instituto de Astrof\'isica de Canarias, The Johns Hopkins University, 
Kavli Institute for the Physics and Mathematics of the Universe (IPMU) / 
University of Tokyo, the Korean Participation Group, Lawrence Berkeley National Laboratory, 
Leibniz Institut f\"ur Astrophysik Potsdam (AIP),  
Max-Planck-Institut f\"ur Astronomie (MPIA Heidelberg), 
Max-Planck-Institut f\"ur Astrophysik (MPA Garching), 
Max-Planck-Institut f\"ur Extraterrestrische Physik (MPE), 
National Astronomical Observatories of China, New Mexico State University, 
New York University, University of Notre Dame, 
Observat\'ario Nacional / MCTI, The Ohio State University, 
Pennsylvania State University, Shanghai Astronomical Observatory, 
United Kingdom Participation Group,
Universidad Nacional Aut\'onoma de M\'exico, University of Arizona, 
University of Colorado Boulder, University of Oxford, University of Portsmouth, 
University of Utah, University of Virginia, University of Washington, University of Wisconsin, 
Vanderbilt University, and Yale University.

This work has made use of data from the European Space Agency (ESA) mission
{\it Gaia} (\url{https://www.cosmos.esa.int/gaia}), processed by the {\it Gaia}
Data Processing and Analysis Consortium (DPAC,
\url{https://www.cosmos.esa.int/web/gaia/dpac/consortium}). Funding for the DPAC
has been provided by national institutions, in particular the institutions
participating in the {\it Gaia} Multilateral Agreement.

\facilities{
	SDSS-IV (APOGEE),
	\gaia}

\software{
	\code{Astropy} \citep{astropy,astropy2}, \,
	\code{matplotlib} \citep{Hunter:2007}, \,
	\code{numpy} \citep{vanderWalt:2011}, \,
	\code{scipy} \citep{Jones:2001}, \,
	\code{Topcat} \citep{Taylor2005}, \,
	\code{The Cannon} \citep{Ness:2015}}


\begin{figure*}
	\begin{center}
	\includegraphics[width=16cm]{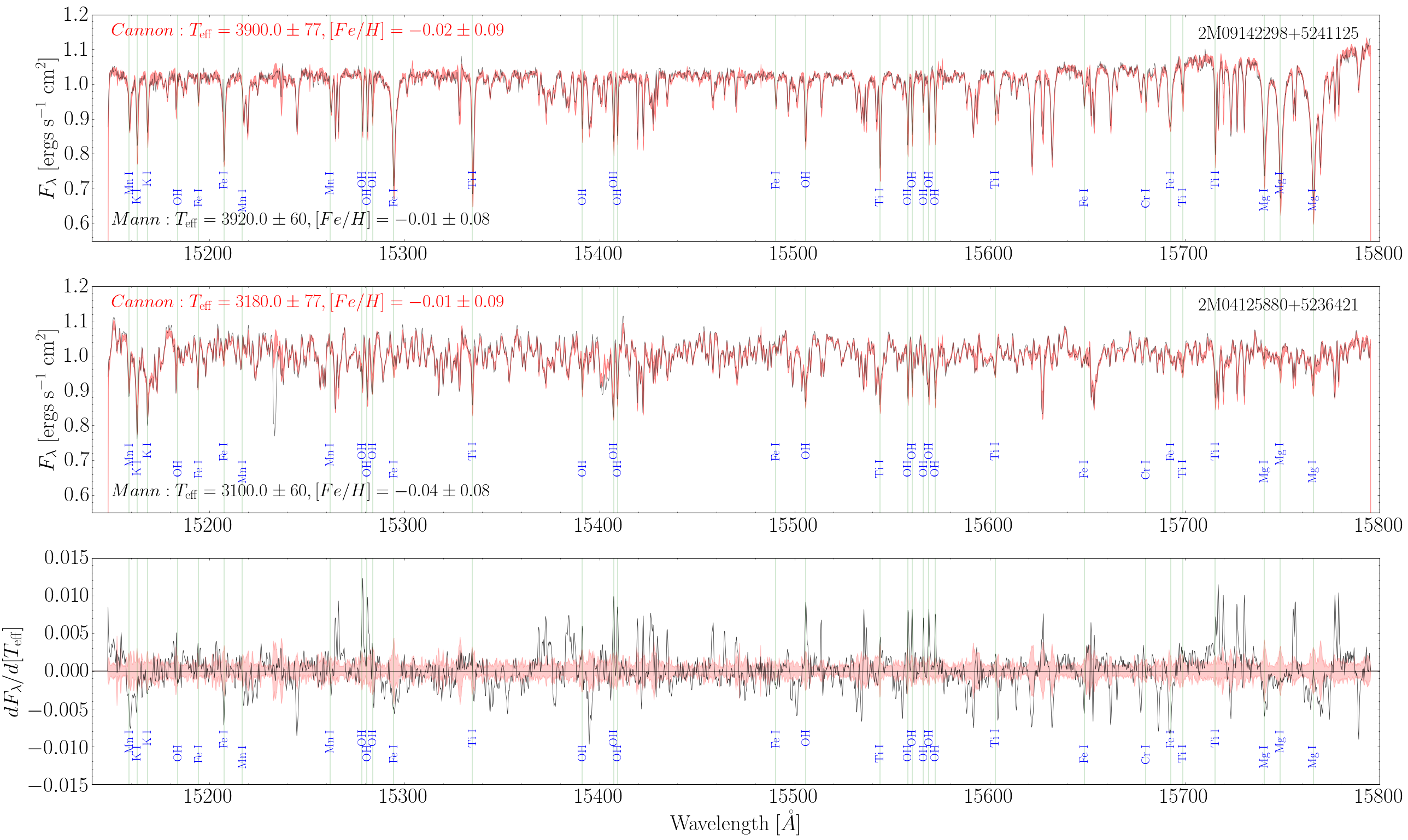} 
	\end{center}
\end{figure*}

\begin{figure*}
	\begin{center}
	\includegraphics[width=16cm]{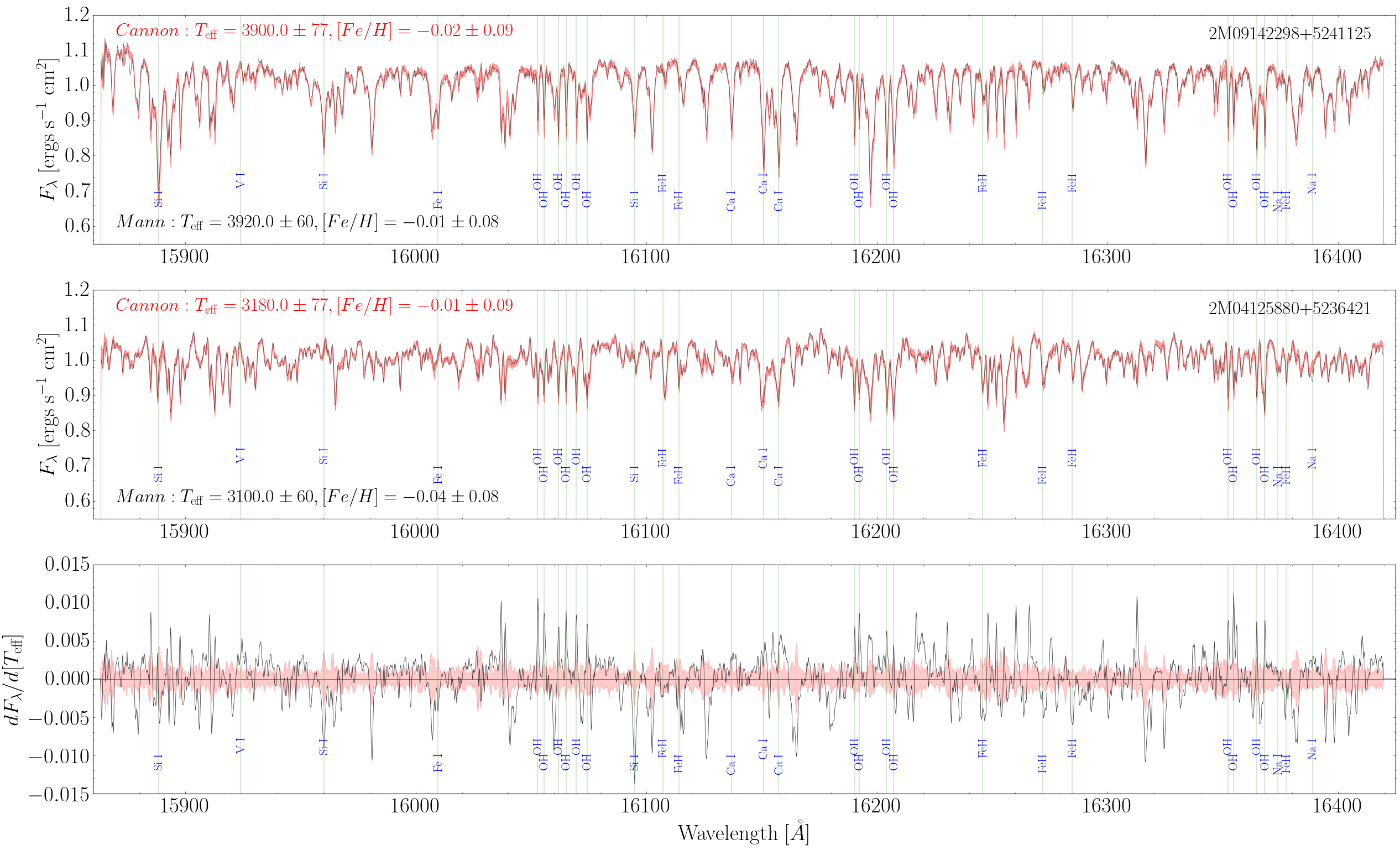}
	\end{center}
\end{figure*}

\begin{figure*}
	\begin{center}
	\includegraphics[width=16cm]{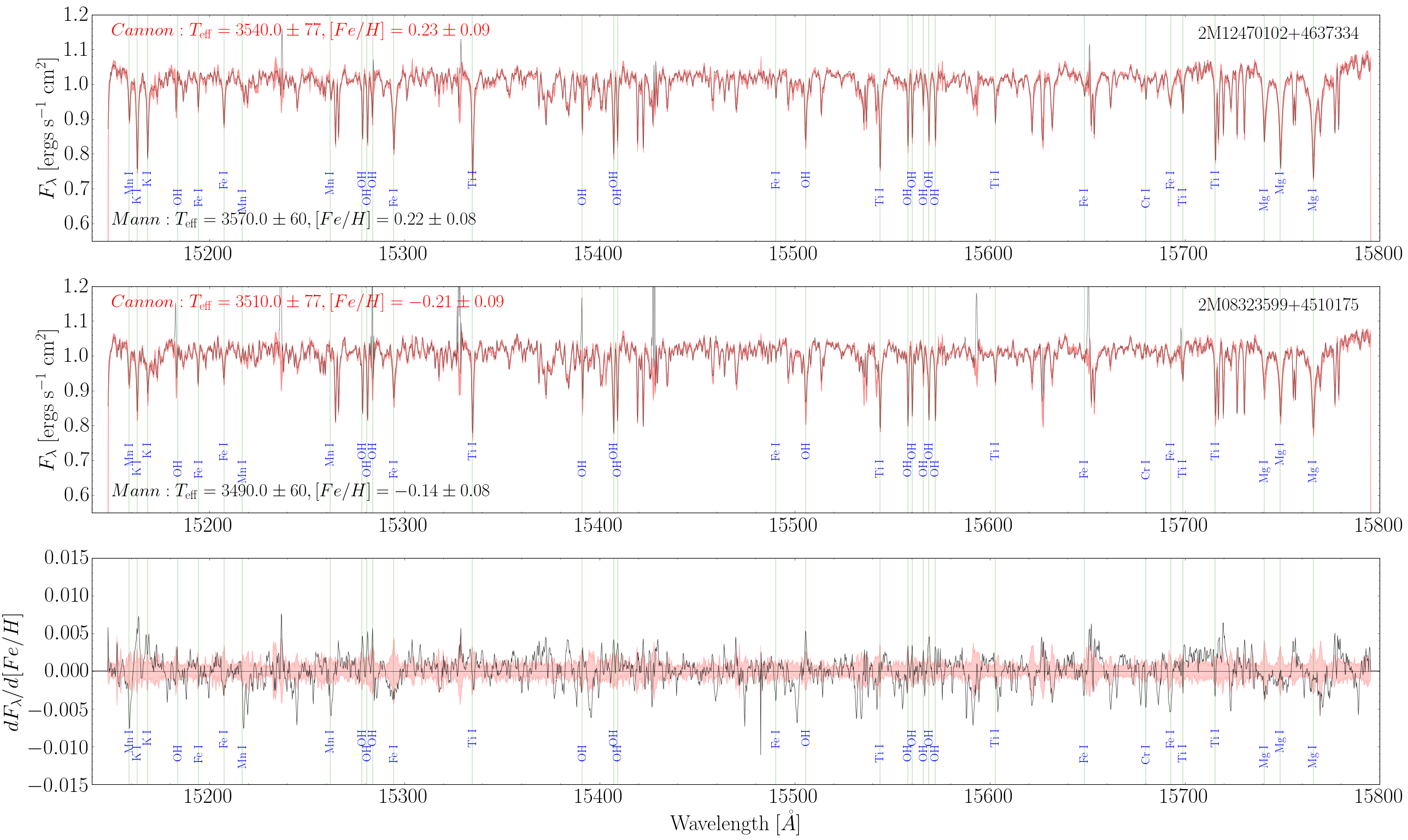}
	\end{center}
\end{figure*}

\begin{figure*}
	\begin{center}
	\includegraphics[width=16cm]{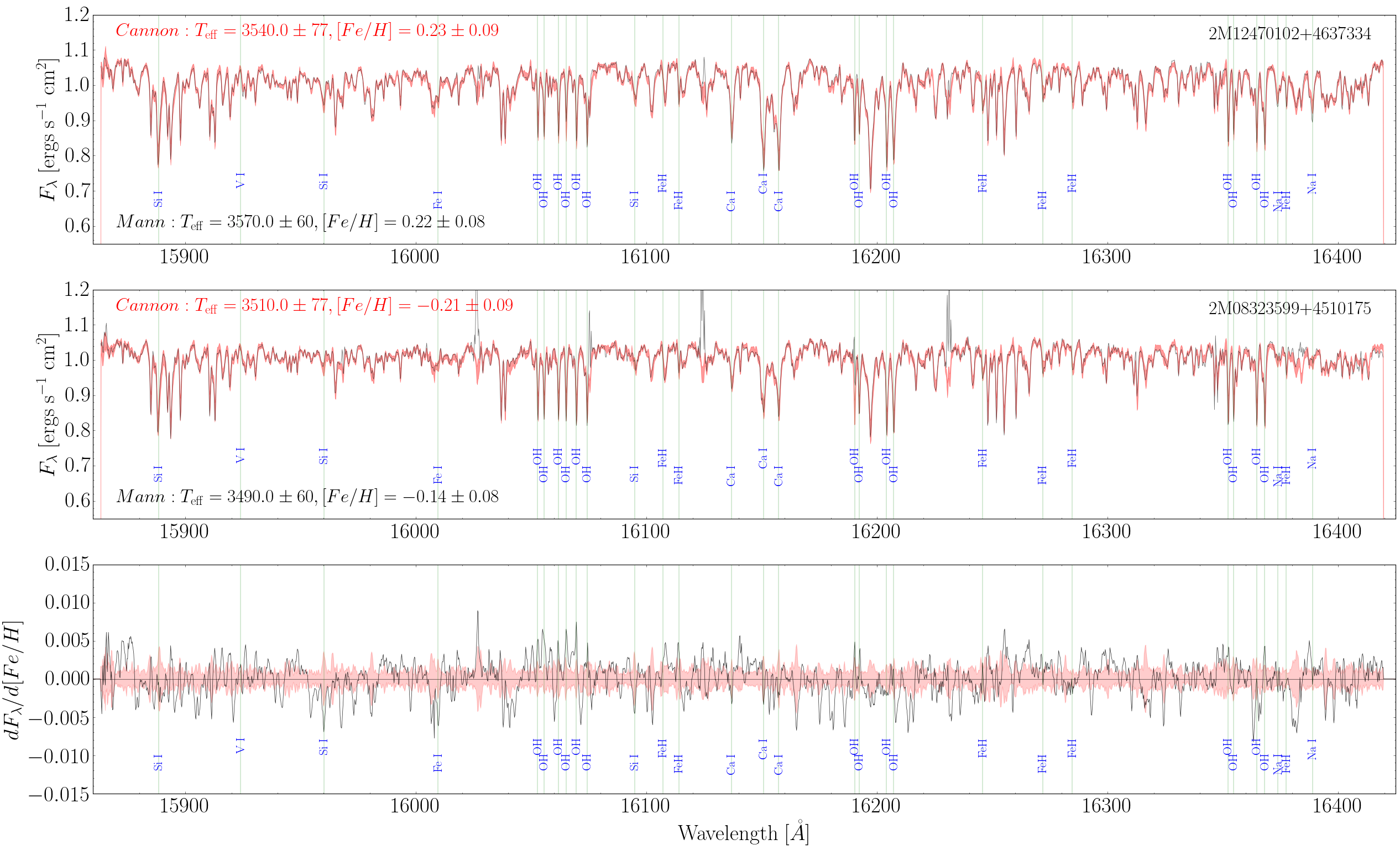}
	\end{center}
\end{figure*}

\newpage

\begin{deluxetable*}{rcl}
\tablecaption{Test results table header description.}
\tablehead{\colhead{Column} & \colhead{Unit} & \colhead{Description}}
	\startdata
	{\tt APOGEE$\_$ID} 	& 	& APOGEE 2MASS designation \\
	{\tt GAIA$\_$ID} 		&	& \gaia\ identification number \\
	{\tt TEFF} 				& K	& Mann-trained \cannon\ effective temperature	\\
	{\tt FE$\_$H }		& dex	& Mann-trained \cannon\ $\feh$	\\
	{\tt SPT} 				& 	& West-trained \cannon\ spectral subtype (M0$-$M9)	\\
	{\tt TEFF$\_$APOGEE} 				& K	& ASPCAP pipeline effective temperature	\\
	{\tt M$\_$H$\_$APOGEE} 				& dex	& ASPCAP pipeline [M/H]	\\
	{\tt CHI$\_$MANN} 	& 	& $\chi^2$ fit of Mann-trained \cannon\ model	\\
	{\tt CHI$\_$WEST} 	& 	& $\chi^2$ fit of West-trained \cannon\ model	\\
	{\tt J$\_$MAG} 				& mag	& APOGEE $J$ band photometry	\\
	{\tt H$\_$MAG} 				& mag	& APOGEE $H$ band photometry	\\
	{\tt K$\_$MAG} 				& mag	& APOGEE $K$ band photometry	\\
	{\tt BP$\_$MAG }		& mag	& \gaia\ $BP$ band photometry	\\
	{\tt RP$\_$MAG} 		& mag	& \gaia\ $RP$ band photometry	\\
	{\tt G$\_$MAG} 		& mag	& \gaia\ $G$ band photometry	\\
	{\tt J$\_$ABS} 		& mag	& APOGEE $J$ band absolute magnitude	\\
	{\tt H$\_$ABS} 		& mag	& APOGEE $H$ band absolute magnitude	\\
	{\tt K$\_$ABS} 		& mag	& APOGEE $K$ band absolute magnitude	\\
	{\tt G$\_$ABS} 		& mag	& \gaia\ $G$ band absolute magnitude	\\
	{\tt BP$\_$RP} 		& mag	& \gaia\ $BP-RP$ color	\\
	{\tt RA} 				& deg	& APOGEE right ascension angle	\\
	{\tt DEC} 			& deg	& APOGEE declination angle	\\
	{\tt PMRA} 			& mas yr$^{-1}$	& \gaia\ right ascension proper motion	\\
	{\tt PMRA$\_$ERR} 	& mas yr$^{-1}$	& \gaia\ right ascension proper motion uncertainty	\\
	{\tt PMDEC} 			& mas yr$^{-1}$	& \gaia\ declination proper motion	\\
	{\tt PMDEC$\_$ERR} 	& mas yr$^{-1}$	& \gaia\ declination proper motion uncertainty	\\
	{\tt PLX} 			& mas	& \gaia\ parallax	\\
	{\tt PLX$\_$ERR} 		& mas	& \gaia\ parallax uncertainty	\\
	{\tt DIST} 			& kpc	& Distance (1/$\varpi$)	\\
	{\tt RV$\_$APOGEE} 	& km s$^{-1}$	& APOGEE radial velocity	\\
	{\tt RV$\_$APOGEE$\_$ERR} 	& km s$^{-1}$	& APOGEE radial velocity uncertainty	\\
	{\tt RV$\_$GAIA} 				& km s$^{-1}$	& \gaia\ radial velocity	\\
	{\tt RV$\_$GAIA$\_$ERR}		& km s$^{-1}$	& \gaia\ radial velocity uncertainty	\\
	{\tt Vx} 				& km s$^{-1}$	& Cartesian $x$ velocity in Galactocentric coordinates	\\
	{\tt Vy} 				& km s$^{-1}$	& Cartesian $y$ velocity in Galactocentric coordinates	\\
	{\tt Vz} 				& km s$^{-1}$	& Cartesian $z$ velocity in Galactocentric coordinates	\\
	{\tt X} 				& kpc	& Cartesian $x$ position in Galactocentric coordinates	\\
	{\tt Y} 				& kpc	& Cartesian $y$ position in Galactocentric coordinates	\\
	{\tt Z} 				& kpc	& Cartesian $z$ position in Galactocentric coordinates  \\
	{\tt SNR} 				& 	& Signal-to-noise ratio of the APOGEE spectrum
\enddata
\end{deluxetable*} \label{table:test_results}


\clearpage
\bibliographystyle{aasjournal}
\bibliography{ref}

\end{document}